\documentclass[lettersize,journal]{IEEEtran}
\usepackage{amsmath,amsfonts}
\usepackage{algorithmic}
\usepackage{array}
\usepackage{textcomp}
\usepackage{stfloats}
\usepackage{url}
\usepackage{verbatim}
\usepackage{graphicx}
\usepackage{subfigure}
\usepackage{subcaption}
\usepackage{bbm}

\def\etc{etc. \@}

\def\BibTeX{{\rm B\kern-.05em{\sc i\kern-.025em b}\kern-.08em
    T\kern-.1667em\lower.7ex\hbox{E}\kern-.125emX}}
\usepackage{balance}
\begin{document}
\title{Machine Learning for Evolutionary Graph Theory}
\author{Guoli Yang, Matteo Cavaliere, Mingtao Zhang, Giovanni Masala, Adam Miles, Mengzhu Wang
\thanks{G. Yang, M. Zhang and M. Wang are with Department of Big Data Intelligence, Advanced Institute of Big Data, E-mail: \{guoliyangcn@outlook.com, mingtaozhang@pku.edu.cn, dreamkily@gmail.com\}}
\thanks{M. Cavaliere is with Modena and Reggio Emilia University, E-mail: matteo.cavaliere@yahoo.com}
\thanks{G. Masala is with the University of Kent, E-mail: a.miles@mmu.ac.uk}
\thanks{A. Miles is with Manchester Metropolitan University. E-mail: g.masala@kent.ac.uk}
}

\markboth{IEEE Transactions on Computational Social Systems}%
{How to Use the IEEEtran \LaTeX \ Templates}

\maketitle

\begin{abstract}
The stability of communities — whether biological, social, economic, technological or ecological depends on the balance between cooperation and cheating. While cooperation strengthens communities, selfish individuals, or "cheaters," exploit collective benefits without contributing. If cheaters become too prevalent, they can trigger the collapse of cooperation and of the community, often in an abrupt manner. A key challenge is determining whether the risk of such a collapse can be detected in advance.
To address this, we use a combination of evolutionary graph theory and machine learning to examine how  one can predict the unravel of cooperation on complex networks. By introducing few cheaters into a structured population, we employ machine learning to detect and anticipate the spreading of cheaters and cooperation collapse. 
Using temporal and structural data, the presented results show that prediction accuracy improves with stronger selection strength and larger observation windows, with CNN-Seq-LSTM and Seq-LSTM best performing models. Moreover, the accuracy for the predictions depends crucially on the type of game played between cooperators and cheaters (i.e., accuracy improves when it is more advantageous to defect) and on the community structure. Overall, this work introduces a machine learning approach into detecting abrupt shifts in evolutionary graph theory and offer potential strategies for anticipating and preventing cooperation collapse in complex social networks.
\end{abstract}


\section{Introduction}
\IEEEPARstart{I}{n}many systems, which evolve between different states of equilibrium, early warning signals (EWS) \cite{Scheffer2009early}, such as return rate, autocorrelation, variance, and so on, can be used to discover the possible "normal forms" that lie beyond the tipping point \cite{esd-15-1117-2024,Dakos2019eco}, which is used then to characterize abrupt changes in response to gradual changes in environmental conditions. This has stimulated research in fields ranging from ecology and biology to sociology, economics and technology \cite{Strogatz2015nonlinear, li2023non, Dakos2019eco, PELISSIE2024110429, dakos2015resilience}, with the aim to develop tools helping to identify if a system is approaching a tipping point. Different terms have been used to describe these phenomena, including critical slowing down (CSD), statistical stability indicators, leading indicators, resilience indicators, indicators of transitions, etc.\cite{Dylewsky2023evo}. People also classify early warnings depending on the type of mechanism (CSD, non-CSD), the data used (temporal, spatial, trait, \etc), and the way they are employed (analyzing patterns, network methods, \etc). The study in early warnings \cite{esd-15-1117-2024} has made available tools and approaches that can be adopted to detect dynamics effectively in a variety of fields. Some available software for the estimation of early warnings are also available in R package, MATLAB, and Python.  Recently, machine learning, especially deep learning, for early warning signals \cite{Thomas2021deep}, has been developed to predict the normal form that characterizes the upcoming tipping point, which illustrates how machine learning can be mined to capture the dynamics of complex systems. A general challenge remains to apply them to real-world scientific domains and challenging questions in an automatized manner, avoiding ad-hoc solutions \cite{Drake2020Alternative, esd-15-1117-2024}. In this paper we focus on the conflict between cooperators and cheaters in evolving and organized communities. Conflicts between cooperative and cheating individuals can be observed in a variety of different systems, among non-human animals and humans, in groups of friends, companies, societies and socio-technological systems \cite{axelrod1981evolution, sachs2004evolution, hammerstein2003genetic,WU2023133692, nowak_s06,wang2025graph}. The main issue is that cooperators pay a cost to distribute benefits while cheaters (known as defectors or free-riders) pay no cost and distribute no benefits \cite{nowak_s06}.
In evolutionary populations, cooperators and defectors will compete with each other for the dominant positions; one of the most studied mechanisms to support cooperation include direct and indirect reciprocity, kin selection, network reciprocity, and group selection \cite{nowak_s06}. However, in a variety of scenarios, cheaters can outcome cooperators and spread in the population leading to the collapse of the community \cite{cavaliere2012prosperity, levin2010crossing, axelrod1981evolution}. Therefore, a major challenge is to identify and understand the mechanisms that can facilitate the resilience of cooperation and predict the future breakdown in the early stages \cite{levin2010crossing}. 
In this work, we intend to develop machine learning models based on temporal and structural information to predict the upcoming dominance of cooperation or defection in evolutionary populations.  This has been attempted in the past \cite{cavaliere2016detecting} using ad-hoc measures, manually designed on a specific model, while here we develop a general early warning methodology based on machine learning and apply it to evolutionary graph theory \cite{lieberman2005evolutionary, perc2010coevolutionary,Ohtsuki2006}, a general model to study evolutionary dynamics on structured populations \cite{kuo2024evolutionary, shakarian2012review, ohtsuki2007evolutionary, sharma2023self, perc2010coevolutionary,wang2024smooth}. 
We assume the reader familiar with basics of evolutionary graph theory \cite{lieberman2005evolutionary,wang2024equity,wang2024inter} and machine learning \cite{alpaydin2020introduction}. However, to fix common definitions and notations, we initially review the main concepts of evolutionary graph theory and of the machine learning methodologies behind the presented study. 
We then present early warnings for evolutionary graph theory, based on machine learning methods using more or less constrained data, and analyse them under the influence of key factors driving the evolutionary dynamics, such as selection strength, network structure, and game parameters.    

\section{Methods}
\label{method}

\subsection{Population Games}
\label{population}

The evolutionary dynamics \cite{nowak2004evolutionary, Yang2021Strategically, Yang2020Ranking, Yang2019Identification} in a population (with size of $N$)  consists of two types of competing agents: cooperators and defectors. In the beginning, the population is full of cooperators. When perturbations happen, a fraction of agents will adopt the strategy of defection, and become defectors to compete with the rest cooperators. Therefore, the initial population composition would be like that:  the fraction of cooperators is $1-\eta$ , and the fraction of defectors is $\eta$. 

In the structured population, agents will play games with each neighbour to gain payoffs. This can be summarized in the payoff matrix which represents the game of Prisoner's Dilemma (PD), where $T>R>P>S$. $R$ is the “reward” payoff if both cooperate. $P$ is the “punishment” if both defect. $T$ is the “temptation” as sole defector, and 
$S$ is the “sucker” payoff as sole cooperator. 

\begin{equation}
\Pi = \bordermatrix{~ & C & D \cr
	C & R & S \cr
	D & T & P \cr}
\end{equation}

In detail, for an agent $x$, its payoff $\pi(x)$ is aggregated by interacting with each adjacent opponent $y$, and is defined in the following way. If $x$ plays $C$ and $y$ plays $C$ as well, then $\pi(x) =\pi(y) = R$. If $x$ plays $C$ and $y$ plays $D$, then $\pi(x) = S$, $\pi(y) = T$. If $x$ plays $D$ and $y$ plays $C$, then $\pi(x) = T$, $\pi(y) = S$. If $x$ plays $D$ and $y$ plays $D$ as well, then $\pi(x) =\pi(y) = 0$.    The intuitive interpretation of the payoff matrix is that an agent that plays $C$ (namely the cooperator) provides a benefit to their opponent, paying a cost; an agent who plays $D$ (namely the defector) pays no cost and gives no benefit. In general, we keep $R=1$ and $P=0$ in this study, while $T>1$ and $S<0$.

For an agent, the final payoff at a given step is calculated by aggregating all payoffs from the neighbourhood.

\begin{equation}
    \pi_t(x) = \sum_{y \in \mathcal{N}_x} \Pi(s_t(x), s_t(y))
\end{equation}
where $\mathcal{N}_x$ is the set of neighbours of $x$, $s_t(x)$ and $s_t(y)$ are the strategy adopted by $x$ and $y$ at step $t$.

In general, the {\em fitness} of an agent is given by a constant term (baseline fitness) plus the payoff obtained by interacting with neighbours. We can define the fitness of an agent $x$ at a certain step $t$ as:

\begin{equation}
f_t(x) = 1+w(\pi_t(x)-1)
\end{equation}
where $w$ is the intensity of selection (namely selection strength) \cite{perc2010coevolutionary,Ohtsuki2006}. 

\subsection{Evolutionary Dynamics}
\label{dynamics}

In this study, we adopt a traditional "death-birth" model \cite{Ohtsuki2006} to update the strategies present in the population. The intuition is that the strategy of an agent who does well is more likely to be imitated by others in the neighbourhood (or more likely to be selected to reproduce). This type of dynamics indicates that more successful agents will be more competitive in  evolution.

An {\em update step} of the computational model studied works as follow: at each step, a random node $x$ is selected to be removed, and its neighbours compete to occupy the empty site with probability proportional to their fitness. The probability of adopting the strategy of cooperation will be:

\begin{eqnarray}
    p(s_{t+1}(x) = C) &= & \frac{\sum_{y \in \mathcal{N}_x, s_t(y) = C}f_t(y)}{\sum_{y \in \mathcal{N}_x}f_t(y)}    
\end{eqnarray}

Similarly, the probability of adopting the strategy of defection will be:
\begin{eqnarray}
    p(s_{t+1}(x) = D) &= & \frac{\sum_{y \in \mathcal{N}_x, s_t(y) = D}f_t(y)}{\sum_{y \in \mathcal{N}_x}f_t(y)} 
\end{eqnarray}

\begin{figure}[!htb]
	\setlength{\abovecaptionskip}{0pt}
	\setlength{\belowcaptionskip}{0pt}
	\centering
	\includegraphics[width=0.8\linewidth]{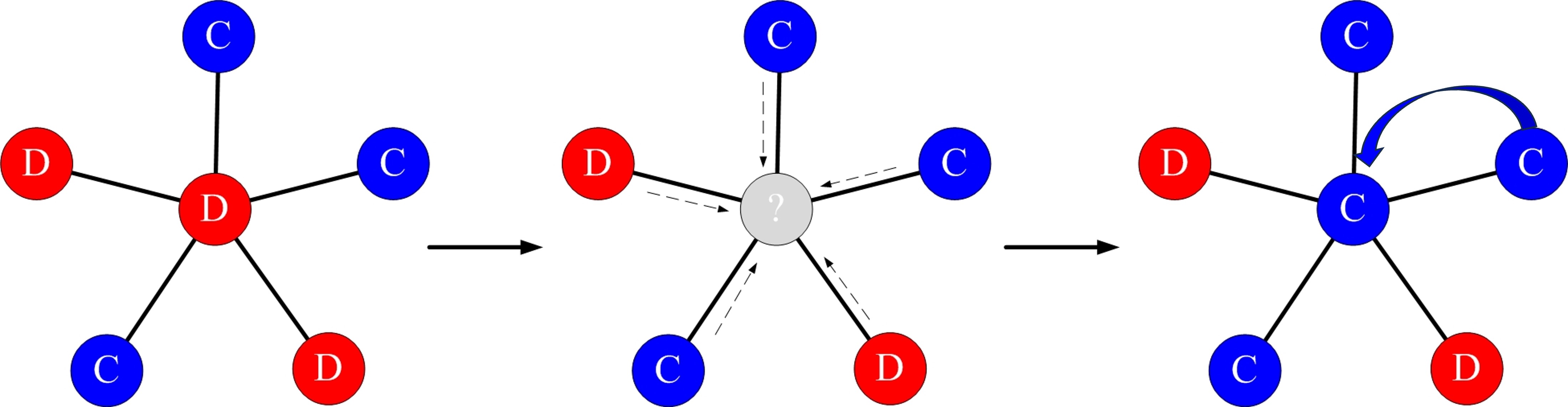}
	\caption{An update step of the model. Each agent occupies the vertex of graph and derives a payoff from the interactions with its neighbours. A random agent is selected to reset its strategy and the new strategy is occupied by one of its neighbours with probability proportional to fitness.}
	\label{fig:model}
\end{figure}

The evolutionary dynamics can then be defined by a discrete sequence of update steps of the type shown in Figure \ref{fig:model}. 

\begin{figure}[!htb]
	\setlength{\abovecaptionskip}{0pt}
	\setlength{\belowcaptionskip}{0pt}
	\centering
	\includegraphics[width=1\linewidth]{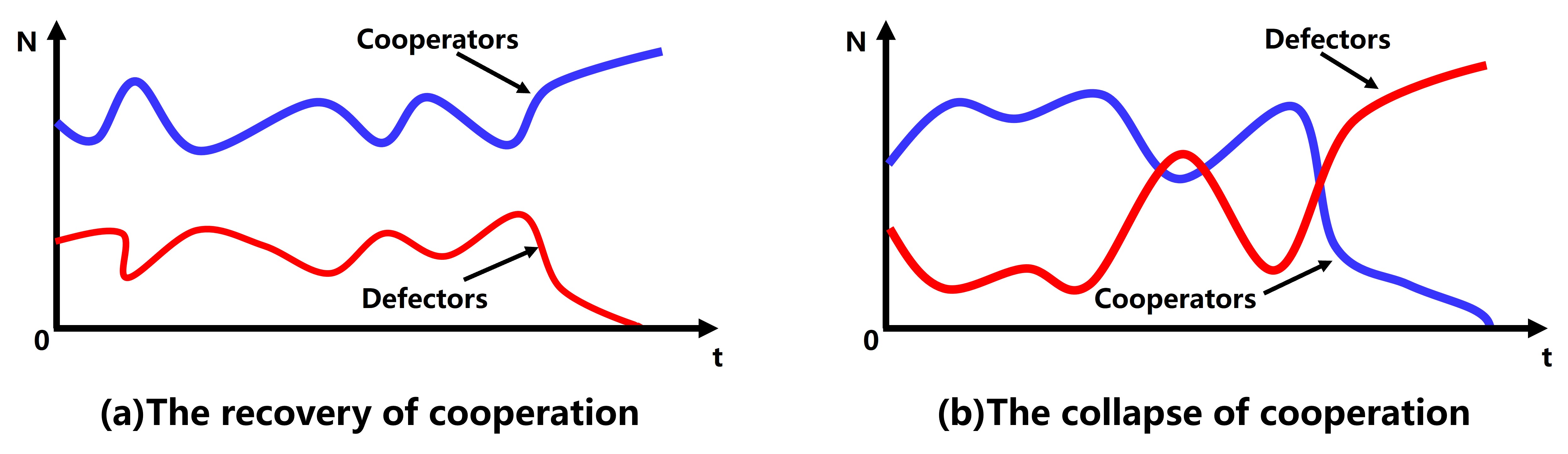}
	\caption{Two frozen states can be obtained in the evolutionary dynamics. One is the recovery of cooperation shown in (a), and one is the collapse of cooperation shown in (b). The recovery time means the duration from the beginning to the equilibrium of recovery, and the collapse time means the duration from the beginning to the equilibrium of collapse. }
	\label{fig:outcomes}
\end{figure}

The evolution of population will continue until one of the two frozen states is reached (Figure \ref{fig:outcomes}) - all nodes are cooperators (i.e., \textit{AllC}) or all nodes are defectors (i.e., \textit{AllD}). For the case of \textit{AllC}, it means that the invasion of defectors fails and the recovery of cooperation is obtained. For the case of \textit{AllD}, it means that the invasion of defectors is successful, and the collapse of cooperation is obtained. In this paper, we are then particulary interested in these two states: collapse (of cooperation) and recovery (of cooperation).

It is known that the strength of selection and the structure of population can influence in different ways the dynamics of the evolutionary game, including the probability of cooperation collapse, and the speed of evolutionary dynamics \cite{cavaliere2016detecting, perc2010coevolutionary, Ohtsuki2006, Yang2021Strategically, Yang2019Identification, Yang2020Ranking, shakarian2012review}. In particular, generally, a weaker strength of selection makes the populations more robust against the invasion of defectors; however, a stronger strength of selection generally facilitates the collapse of cooperation. Furthermore, with the increase of selection strength, both the recovery time and the collapse time will decrease, indicating that higher strength of selection will speed up the evolution of population \cite{cavaliere2016detecting, Yang2021Strategically, Yang2019Identification, Yang2020Ranking}. 

\subsection{Early Warning Prediction}
\label{prediction}

The evolutionary dynamics described above leads to two quite different outcomes, and it requires effective approaches to obtain qualitative information characterizing the temporal and structural sequences. In this way, one can better prepare for the coming situations or take actions to avoid the undesirable state. Considering the significance in anticipating population recovery or collapse, we develop an early warning prediction framework based on machine learning models.

\begin{figure}[!htb]
	\setlength{\abovecaptionskip}{0pt}
	\setlength{\belowcaptionskip}{0pt}
	\centering
	\includegraphics[width=0.7\linewidth]{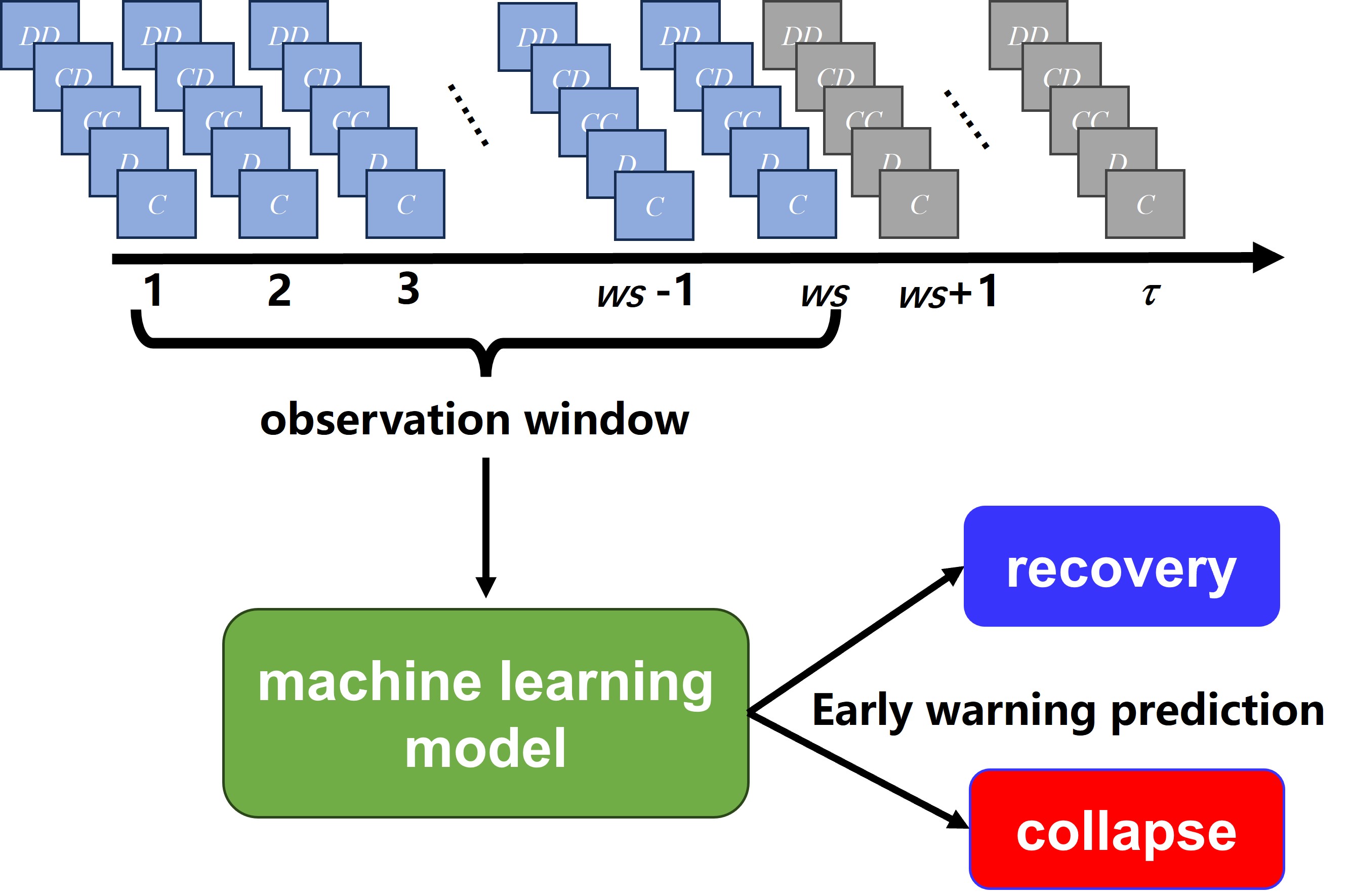}
	\caption{The early warning prediction framework based on machine learning models. For a fixed observation window with size $ws$, the varying numbers of nodes and edges are used for machine learning model training to predict the oncoming recovery of cooperation or collapse of cooperation. Bigger value of $ws$ indicates more temporal and structural information are collected to train the machine learning model.}
	\label{fig:prediction}
\end{figure}

The early warning prediction framework contains three components, as shown in Figure \ref{fig:prediction}. 
\begin{itemize}
    \item Data collection:  for a given observation window, the temporal and structural sequences (including the number of cooperators ($\#C$), the number of defectors ($\#D$), the number of mutual cooperative links ($\#CC$), the number of successful exploitative links ($\#CD$), and the number of mutual defective links ($\#DD$) ) from the beginning of evolution to the end of the observation window (with a length of $ws$) are collected as training set.
    \item Model training: using the temporal and structural sequences and the respective outcomes (either \textit{AllC} or \textit{AllD}), some machine learning models, which can deal with the sequential data well, are trained.
    \item Performance evaluation: the trained models are adopted to perform the prediction of a set of new evolutionary populations.
\end{itemize}

To find the normal patterns lied beyond the sequential and structural evolutionary populations, we adopt some deep learning models to extract the features that appear in the evolutionary sequences (i.e., the dynamical number of nodes, edges, or motifs). Those models can tackle different features in a long time series, constructing a reasonable mapping from the features mentioned above to the probability of the series being true or false (i.e., the recovery or collapse of cooperation). As a result, those approaches excel at time series mining, structural pattern recognition and early warning prediction. 

To make it more feasible and effective for early warning prediction, we design a set of reconfigurable flexible assembly units to construct the temporal and structural machine learning models. The following are descriptions of the architecture of the models, which is composed of LSTM-Layer, CNN-Layer, Self-Attention-Layer, and Fully-Connected-Layer.

(1) LSTM-Layer~\cite{LSTM}, short for Long Short Term Memory, is a prevalent form of Recurrent Neural Network (RNN). By incorporating a cell state $C_t$ to accumulate information of the previous step and the current input $x_t$ ($C_{t-1}$ as the previous cell state, and $h_{t-1}$ as the previous hidden state), LSTM layer effectively captures long-term dependencies between the sequence data. The process can be formalized as follows: $h_{t-1}$ and $x_t$ are used by the forget, input and output gate to generate the decaying rate $f_t$, where the input rate is $i_t$ and the output rate is $o_t$ respectively. The gates can be calculated by $(f/i/o)_t=\sigma(W_{(f/i/o)}\cdot[h_{t-1},x_t]+b_{(f/i/o)})$. The current cell state $C_t$ is formalized as $C_t=f_t\circ C_{t-1}+i_t\circ \Tilde{C}_{t}$, where $\circ$ donates element-wise multiplication, $f_t\circ C_{t-1}$ represents the retained information, and $i_t\circ \Tilde{C}_{t}$ represents the current input information. $\Tilde{C}_{t}$ is computed as $\Tilde{C}_t=tanh(W_c\cdot[h_{t-1},x_t]+b_t)$. Finally, the hidden state of step $t$ is represented as $h_t=o_t\circ tanh(C_t)$.

(2) CNN-Layer~\cite{CNN}, an abbreviation for Convolutional Neural Network, is a widely adopted structure. It excels at extracting features from the input by sliding a small window across the input and using convolution operation to aggregate the local information. The extraction process relies on the convolution operation, which can be formalized as $(f*g)(x)=\int_{-\infty}^{\infty}f(u)g(x-u)du$ where $x$ is a continuous variable and $g$ is the learnable convolutional kernel. With $g$ serving as a learnable parameter, the kernel can be flipped and the operation can be simplified to $Conv_g(X)=\sum_{u,v}X_{u,v}g_{u,v}$ for the convolutional layer with 2-D kernel. In the scenario of sequential data, CNNs extract features by aggregating contextual values, but overlook the sequential order information of the input. The specific type of CNN layer is determined by the kernel dimension; for instance, "Conv-1d" aggregates data feature along a single dimension (with convolution applied independently to each position), whereas "Conv-2d" further aggregates data of different positions. 

(3) Self-Attention-Layer~\cite{Transformer} is a commonly utilized mechanism for the tasks involving sequential data that exhibits contextual dependencies. It transforms the input by capturing non-local relationships among the input data, with weights computed using a scaled dot-product operation donated as: $\emph{attention}(q,k,v)=\emph{softmax}(\frac{q^Tk}{\sqrt{d}})v$, where $ \emph{softmax}(z_i)=\frac{exp(z_i)}{\sum_jexp(z_j)}$. The query ($q$) , key ($k$), and value vectors ($v$) , which are all elements of $\mathbbm{R}^d$, are typically calculated by $(q/k/v)=W_{(q/k/v)}x$ respectively, where $W_{(q/k/v)}$ are all learned weight matrices. Ultimately, the attention layer applied to an input $X\in \mathbbm{R}^{ws\times d}$ can be expressed as $\emph{attention}(X)=\emph{softmax}(\frac{(XW_q)(XW_k)^T}{\sqrt{d}})XW_v$. 

(4) Fully-Connected-Layer is a well-known layer, which can be represented mathematically by the equation $y=\sigma(Wx+b)$. Initially, the layer performs a linear transformation on the input features, followed by the application of the activation function $\sigma$, which introduces non-linearity. 

\begin{figure}[!htb]
	\centering
	\subfigure[Transformer]
	{		\includegraphics[height=0.4\textwidth, width=0.22\textwidth]{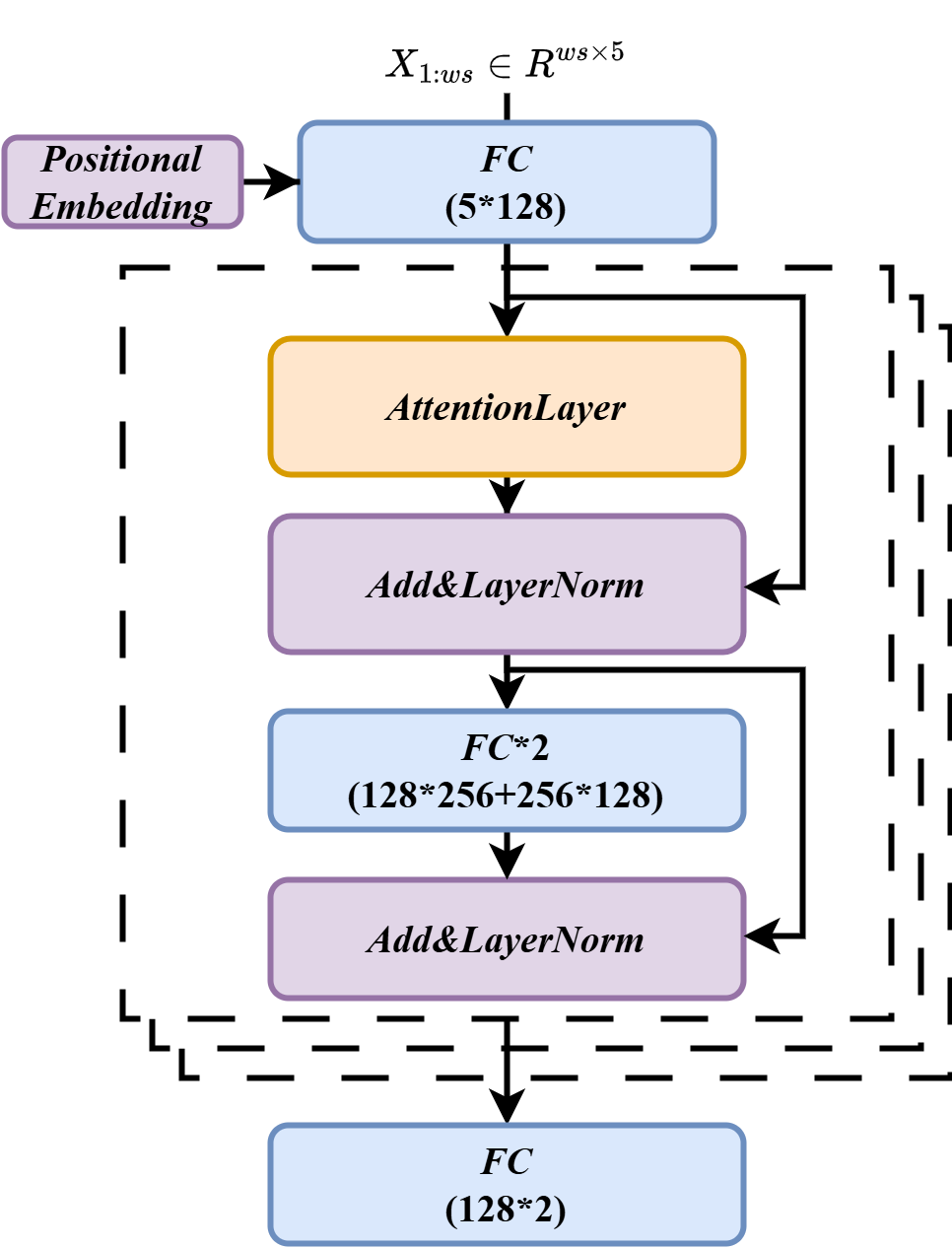}	
	}	
	\caption{The structures of $5$ deep learning models constructed by reconfigurable flexible assembly units.}
	\label{fig:mlmodel}
\end{figure}

For brevity, we omit the detailed explanations of certain secondary layers, such as LayerNorm (which standardizes the input to a distribution with a mean of $0$ and a deviation of $1$, with an optional linear transformation) and Embedding (which assigns each position with a unique learnable embedding) that are mainly utilized in Transformer model~\cite{Transformer} , and Maxpooling (which selects the maximum value from specific positions) that are only utilized in CNN model~\cite{CNN}. By combining these units in various ways, different machine learning models can be constructed, as illustrated in Figure \ref{fig:mlmodel}.

\begin{figure}[!htb]
	\centering
	\subfigure[Seq-LSTM]
	{		\includegraphics[height=0.4\textwidth, width=0.14\textwidth]{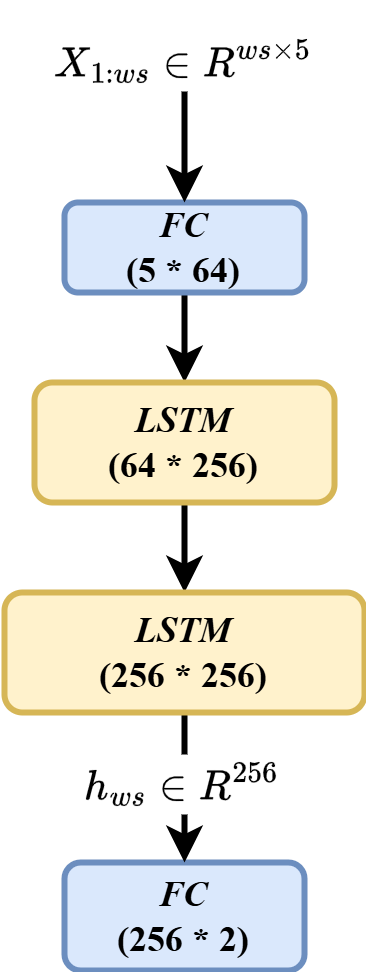}
	}
       \subfigure[CNN-Seq-LSTM]
	{		\includegraphics[height=0.4\textwidth, width=0.17\textwidth]{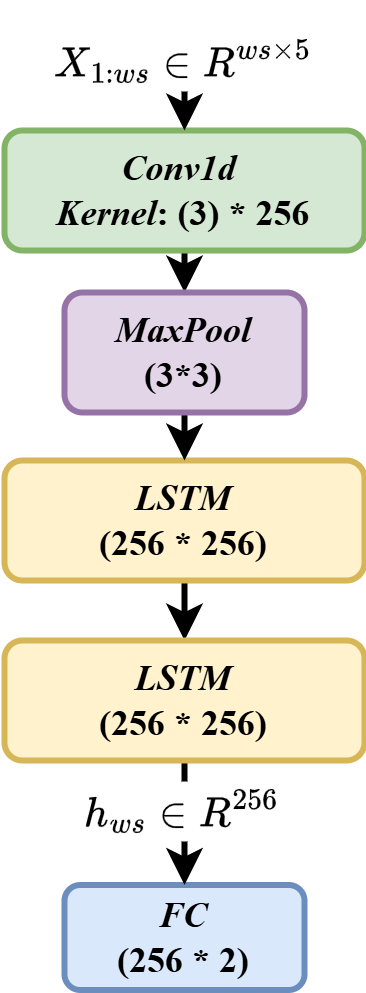}	
	}
	\subfigure[CNN-LSTM]
	{		\includegraphics[height=0.4\textwidth, width=0.14\textwidth]{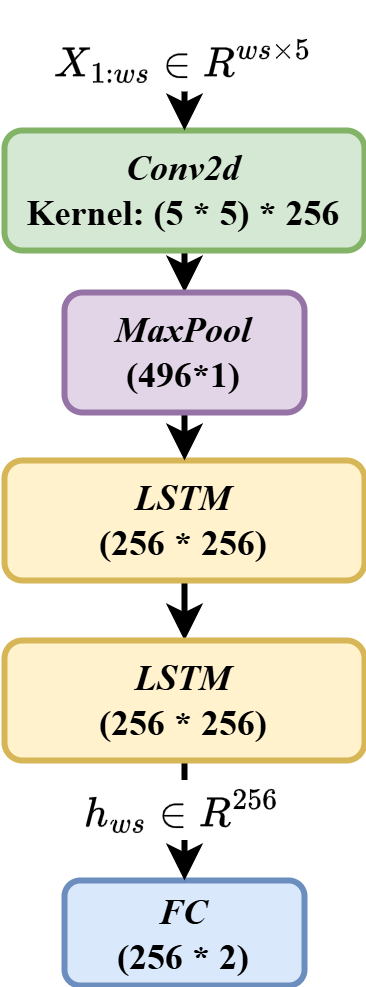}	
	}	
        \subfigure[Text-CNN]
	{		\includegraphics[height=0.4\textwidth, width=0.24\textwidth]{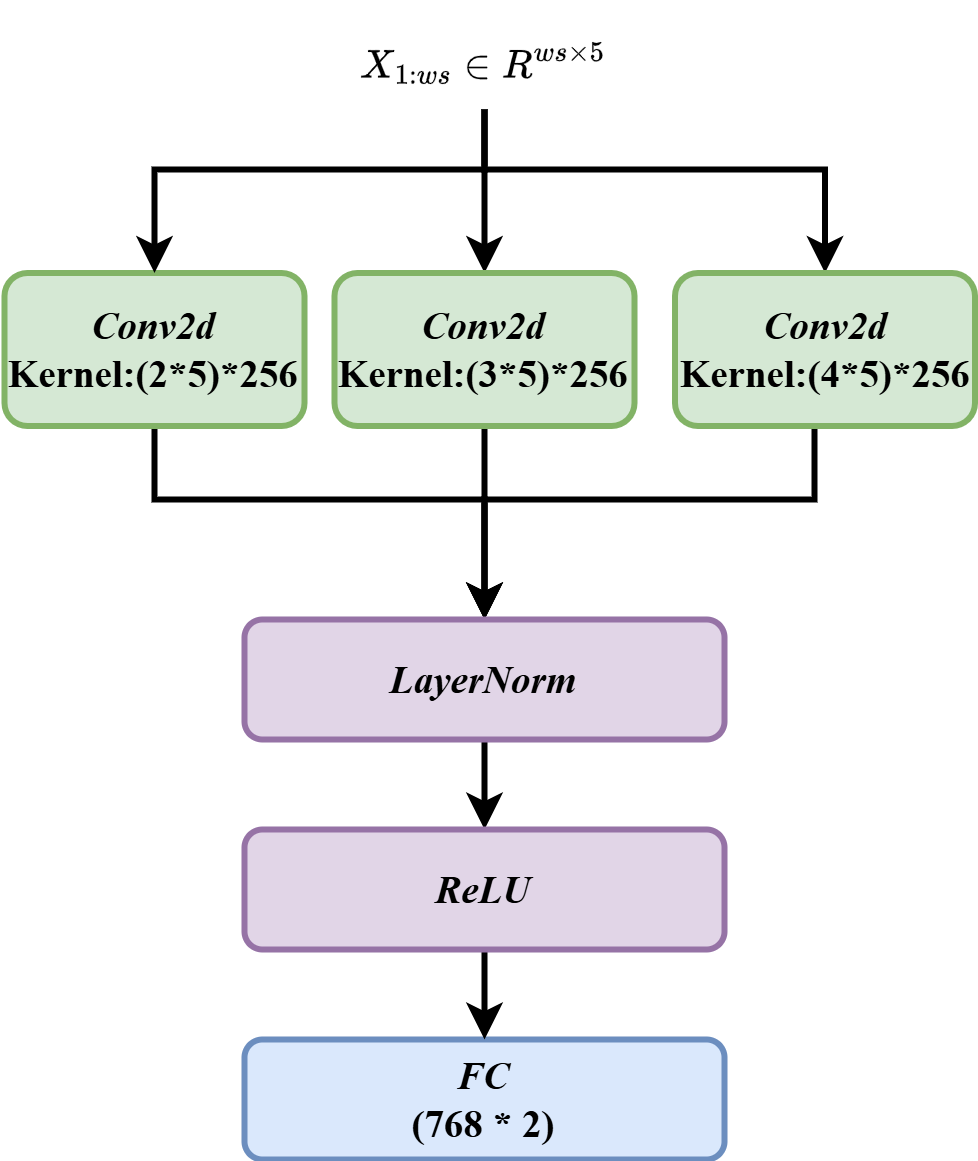}	
	}
\end{figure}

In this study, we consider the following $5$ models as representatives to predict the evolutionary dynamics based on temporal and structural information.

\textbf{Seq-LSTM}: the input data traverse two layers of LSTM~\cite{LSTM} to capture the sequential hidden states in the evolution of the population. Subsequently, a fully connected layer maps these hidden states to the probability of the sequence being classified into a specific label.

\textbf{CNN-Seq-LSTM}: this model employs a CNN layer~\cite{CNN} to aggregate the input at each position into representations within the sequences. Specifically, Conv1d is utilized as the CNN layer. Following max pooling, a structure similar to Seq-LSTM is implemented, consisting of two LSTM layers and a fully connected layer.

\textbf{CNN-LSTM}: this framework is analogous to CNN-Seq-LSTM, with the sole distinction being the substitution of the convolutional layer with "Conv2d", allowing the processing of multiple sequence positions simultaneously.

\textbf{Text-CNN}: this model incorporates three convolutional layers with distinct kernel sizes, enabling it to capture local input information on various scales. The outputs of these kernels are concatenated into a flattened vector and subsequently projected through a fully-connected layer to produce the final output.

\textbf{Transformer}: in this model~\cite{Transformer}, we initially project the features at each position into a high-dimensional representation. We then augment this representation with positional encoding and pass it through the standard Transformer-Encoder architecture (which consists of $3$ layers of modules, as depicted in Figure~\ref{fig:mlmodel}). Finally, the outputs are fed through a fully-connected layer.

\begin{figure}[!htb]
	\setlength{\abovecaptionskip}{0pt}
	\setlength{\belowcaptionskip}{0pt}
	\centering
	\subfigure[recovery case]
	{
        \includegraphics[width=0.48\linewidth]{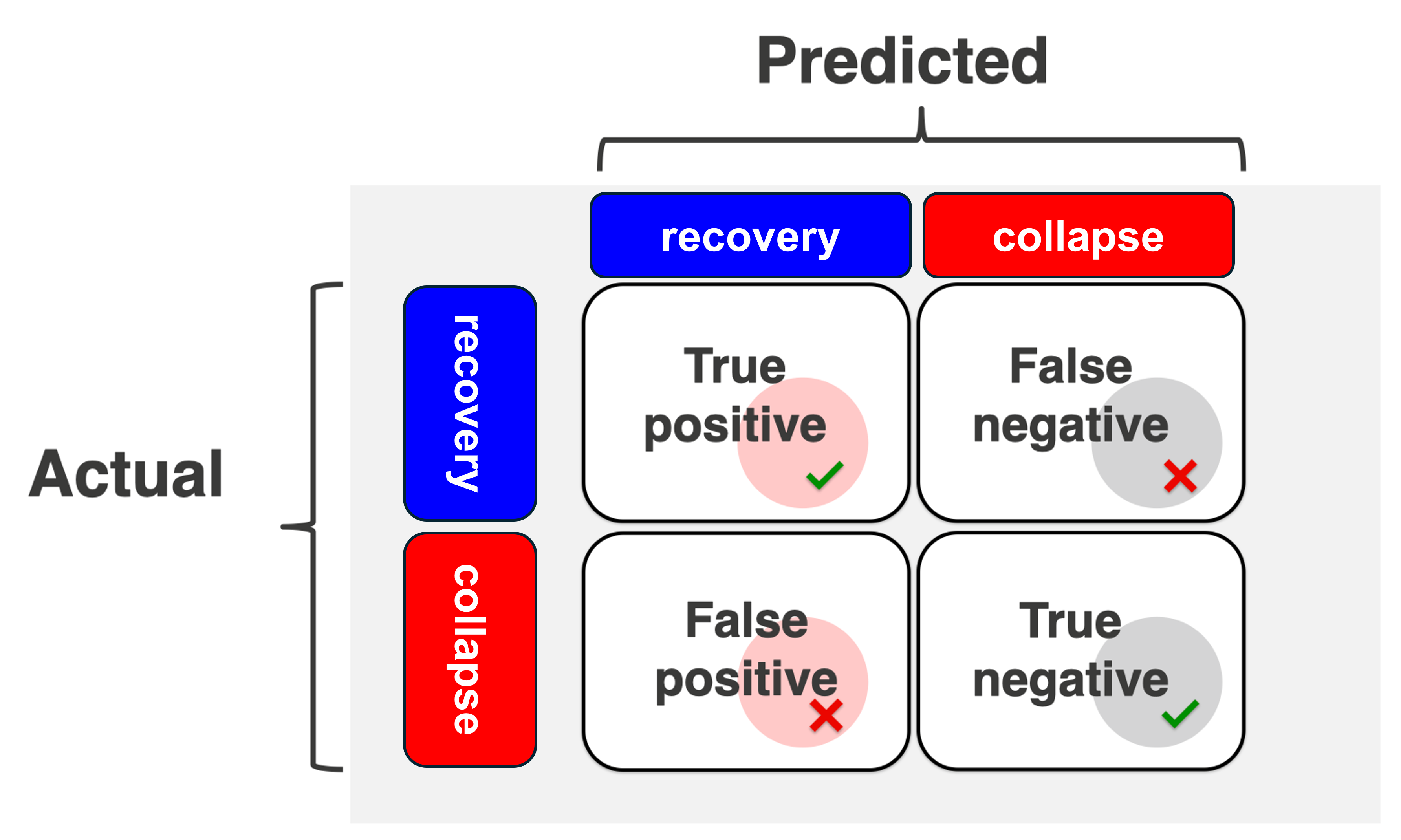}
        }
    \subfigure[collapse case]
	{
        \includegraphics[width=0.48\linewidth]{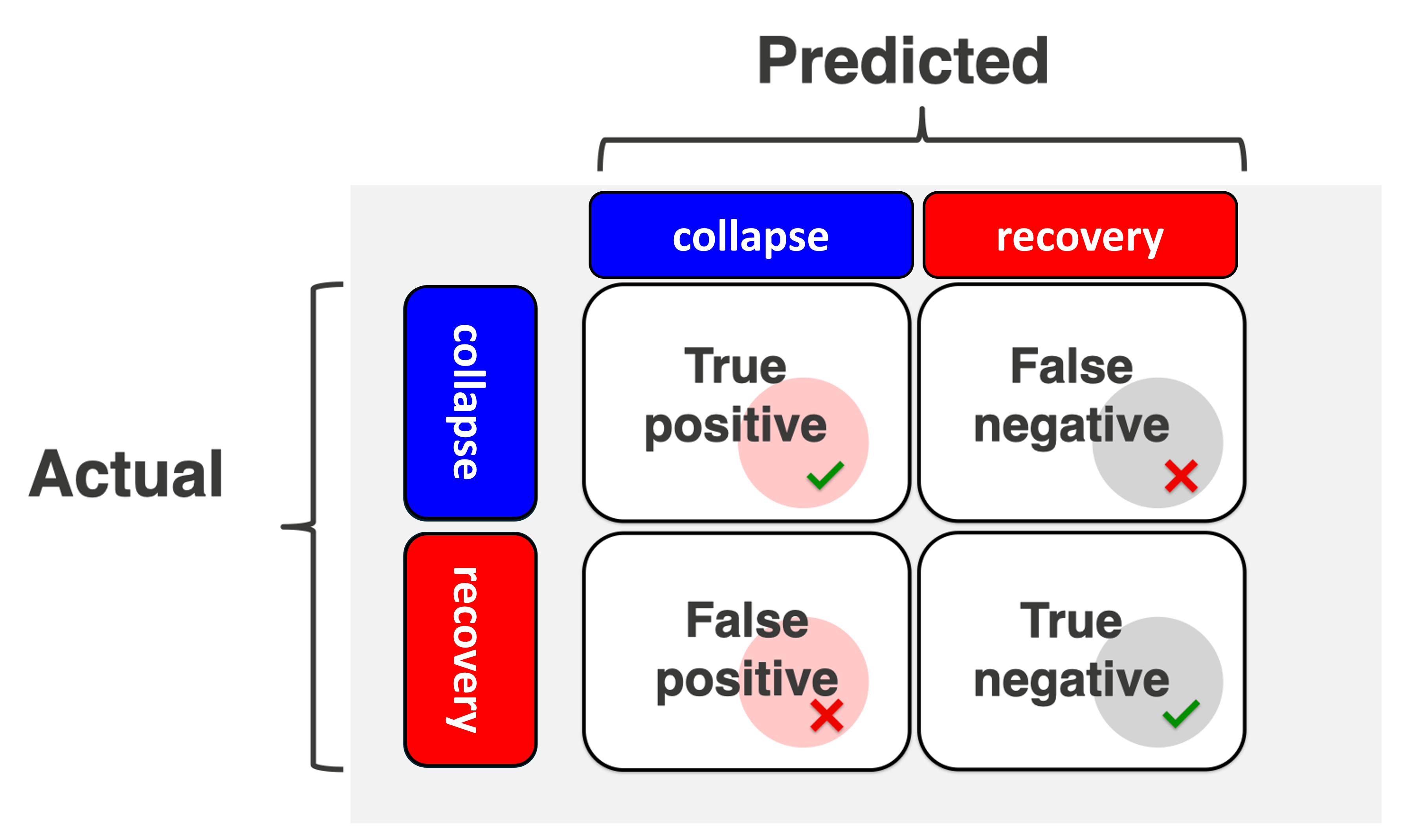}
        }
	\caption{The confusion matrix illustrates the correct predictions and model errors.   Correct predictions include true positives and true negatives, and model errors include false positives and false negatives. For the case of recovery prediction in (a), a true positive means both actual and predicted outcomes are recovery. For the case of collapse prediction in (b), a true positive means both actual and predicted outcomes are collapse.}
	\label{fig:confusion}
\end{figure}

When targeting a certain label (such as the case of \textit{recovery} in the example of Figure \ref{fig:confusion}(a)), there can be $4$ types of classification results, which can be denoted as: TP (True Positive, classifies the evolutionary sequences samples with the targeted label correctly, namely actual \textit{AllC} and predicted  \textit{AllC} ), FP (False Positive, classifies the samples with the other label to be the targeted label, namely actual \textit{AllD} and predicted \textit{AllC}), TN (True Negative, classifies the samples with the other label not be the targeted one, namely actual \textit{AllD} and predicted \textit{AllD} ), FN (False Negative, classifies the samples with the targeted label to be the other types, namely actual \textit{AllC} and predicted  \textit{AllD}). Using the confusion matrix, we can visualize the $4$ different outcomes for the prediction of early warnings.

The following performance metrics are used to evaluate how well these machine learning models can perform in the early warning prediction for evolutionary populations.

\textbf{Precision} (P) measures how often the machine learning model correctly predicts the recovery of cooperation (the positive case in this example shown in Figure \ref{fig:confusion}), which is calculated by:
\begin{equation}
   P = \frac{\#TP}{\#TP+\#FP}
\end{equation}

\textbf{Recall} (R) measures how often a machine learning model correctly identifies the recovery of cooperation from all actual recovery cases in the dataset, which is calculated by:
\begin{equation}
    R = \frac{\#TP}{\#TP+\#FN}
\end{equation}

\textbf{F1 score} (F1) takes both precision and recall into account to make a balance of the two, and it is calculated by:
\begin{equation}
    F1 = 2*\frac{P\times R}{P+R} = \frac{\#TP}{\#TP+0.5(\#FP+\#FN)}
\end{equation}

\textbf{Accuracy} (ACC) measures how often a machine learning model correctly predicts the outcomes (both the case of recovery and the case of collapse), and it is calculated by:
\begin{equation}
   ACC = \frac{\#TP+\#TN}{\#TP+\#FP+\#TN+\#FN}
\end{equation}

Notably, ACC is the metric that is indifferent from the prediction of recovery to the prediction of collapse. The rest $3$ differ from one to the other.

\section{Results}

\label{results}

To verify how successful those machine learning models can predict the oncoming recovery or collapse of cooperation, a series of identical and independent simulations are conducted. In particular, we will study the influences of games, networks, selection strength, and observation windows on the performance of the predictions.

\subsection{Higher strength of selection requires larger size of observation window}

\begin{figure}[!htb]
	\setlength{\abovecaptionskip}{0pt}
	\setlength{\belowcaptionskip}{0pt}
	\centering
	\subfigure[$w=0.001$]
	{		\includegraphics[width=0.48\linewidth]{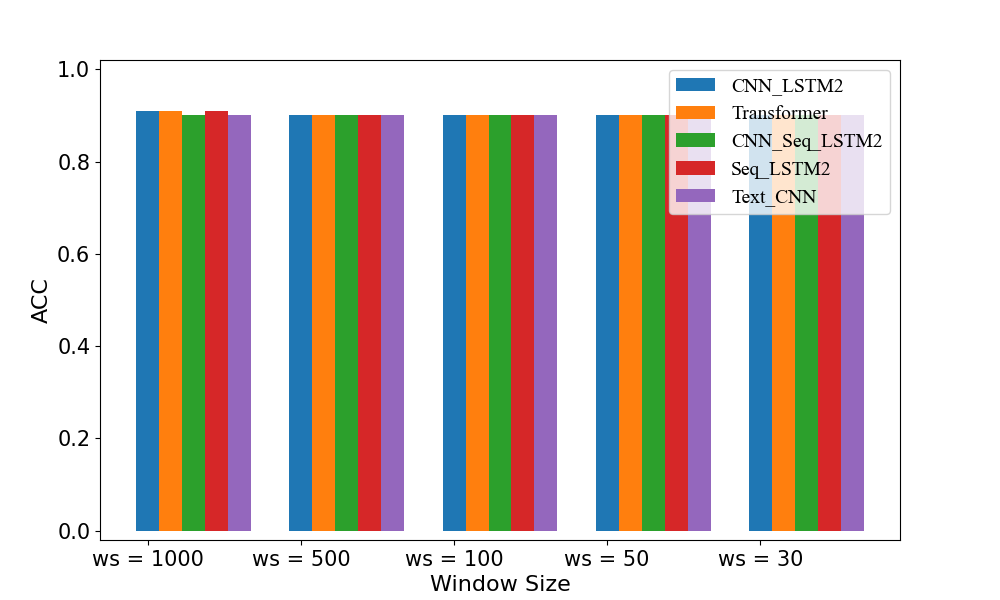}
	}
         \hskip -8pt
	\subfigure[$w=0.005$]
	{		\includegraphics[width=0.48\linewidth]{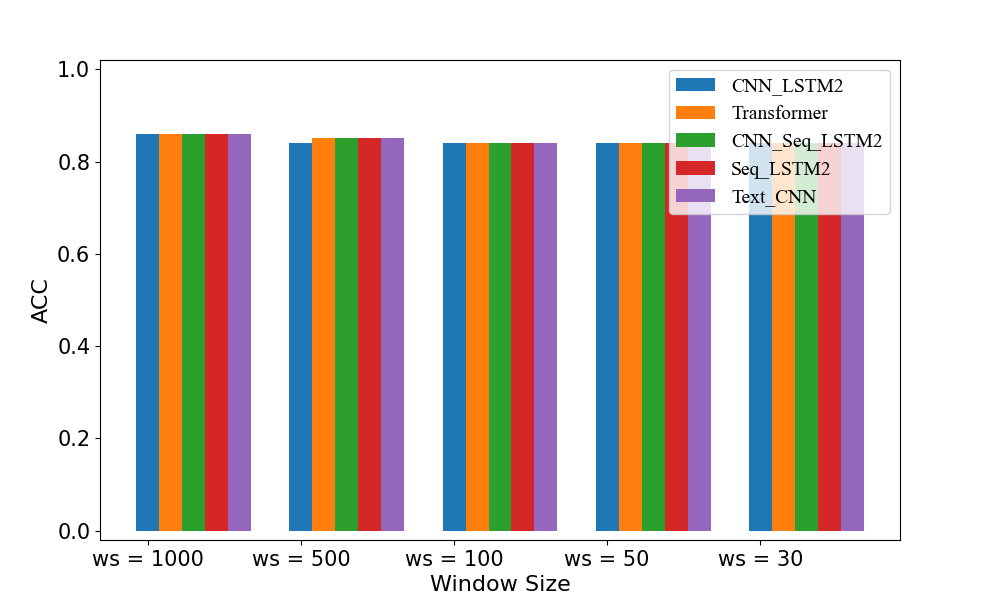}	
	}
 \hskip -8pt
	\subfigure[$w=0.01$]
	{		\includegraphics[width=0.48\linewidth]{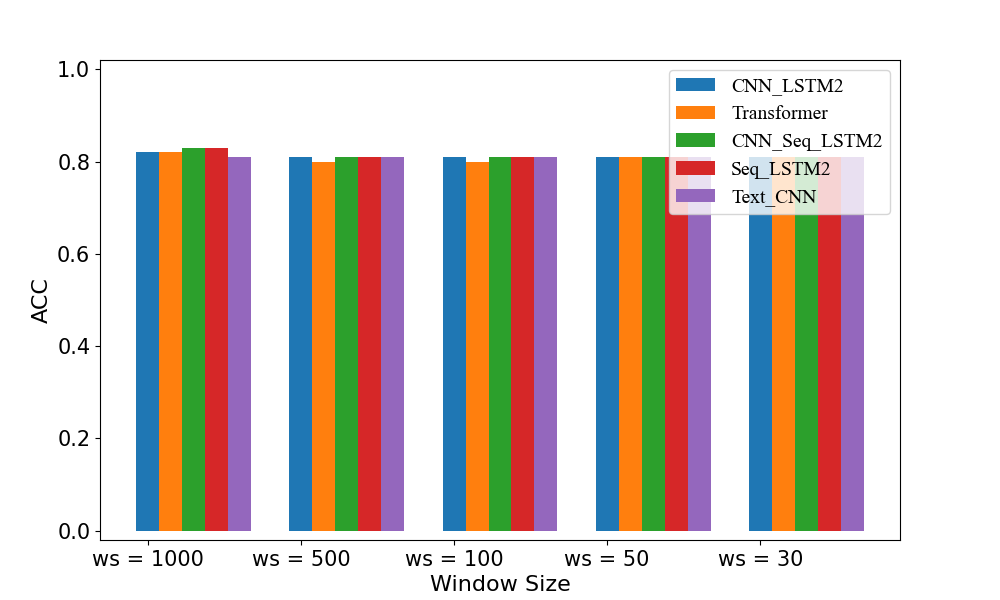}	
	}	
 \hskip -8pt
        \subfigure[$w=0.05$]
	{		\includegraphics[width=0.48\linewidth]{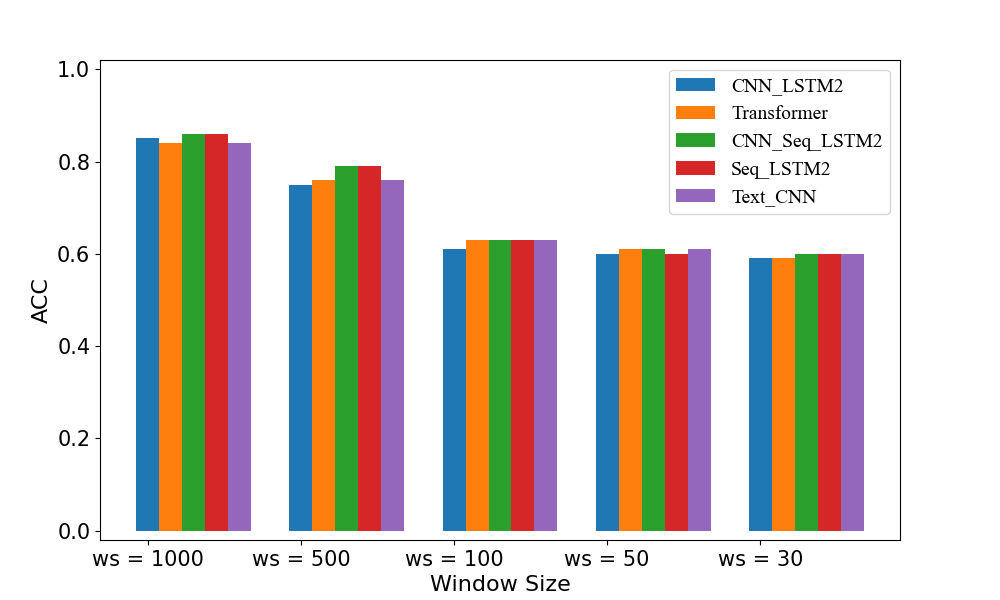}	
	}
 \hskip -8pt
	\subfigure[$w=0.1$]
	{		\includegraphics[width=0.48\linewidth]{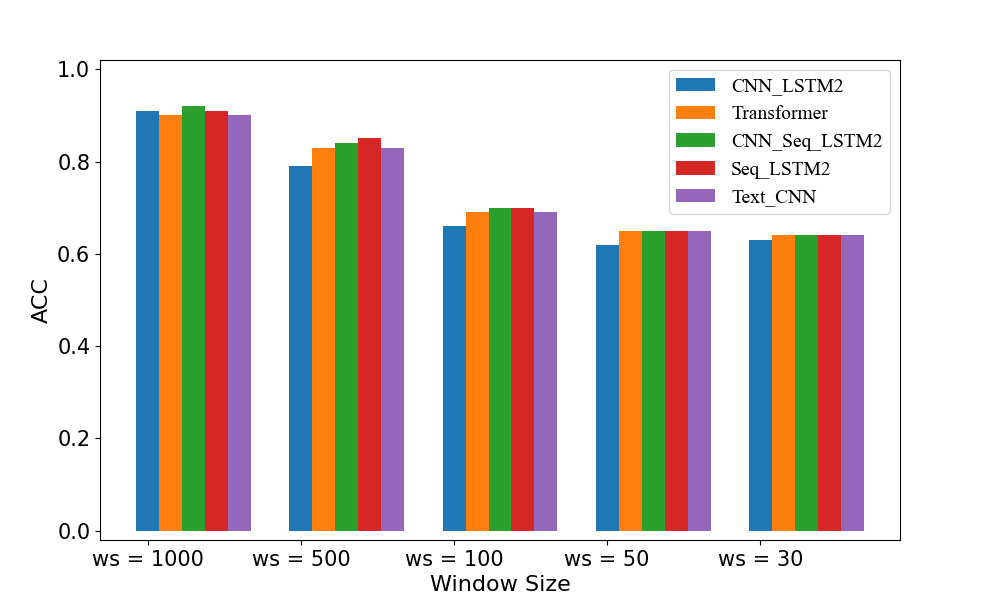}	
	}	 \hskip -8pt
	\subfigure[average]
	{		\includegraphics[width=0.48\linewidth]{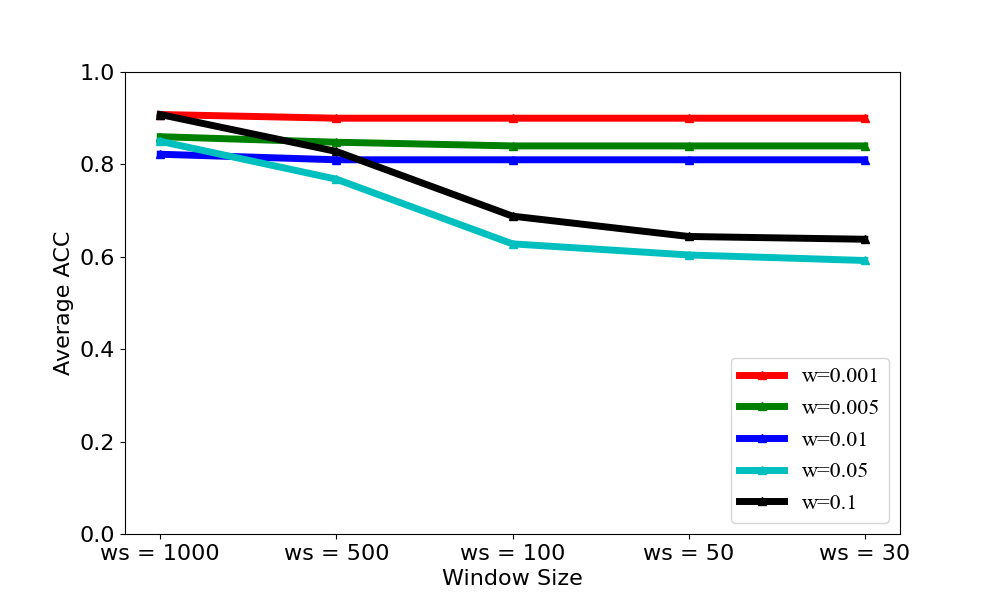}	
	}	
	\caption{The accuracy of early warning prediction is dependent on the strength of  selection  and the size of observation windows. Total $20000$ identical and independent simulations are conducted in a population with $N=100$ agents, where the initial fraction of defectors is $\eta = 10\%$, and that of cooperators is $1-\eta = 90\%$. The game payoff matrix is set to be: $R=1$, $P=0$, $S =-1$, and $T=2$. When the strength of selection $w$ is weak (for example, subfigure (a), (b) and (c)), the prediction performance is independent on the size of observation window $ws$. However, when the strength of selection is strong (for example, subfigure (d) and (e)), larger size of observation windows will bring higher accuracy. Subfigure(f) illustrates the average values of accuracy for varying selections and windows, where a decrease trend is available for small-size windows at strong selection. }
	\label{fig:acc}
\end{figure}

To verify the performance of various machine learning models in early warning prediction, we present the values of \emph{ACC} under different sizes of observation windows and different strengths of selection. It is interesting that all models perform similarly with high performance at weak selection (Figure \ref{fig:acc}(a, b, c)). However, the models of \textbf{CNN-Seq-LSTM} and \textbf{Seq-LSTM} have better performance compared to other methods at strong selection (Figure \ref{fig:acc}(d, e)). This can be caused by the superior ability of LSTMs in sequence modeling and the selection of channel-independent features, both of which are known to be more robust when solving temporal or sequential data. In addition, the size of observation windows will affect the performance of prediction at strong selection (for example $w=0.05$ and $w=0.1$). When the size of observation windows is large (for example $ws = 500$ or $ws=1000$),  the sequences used for machine learning models' training contain enough long time of structural and temporal information, and those models can learn the "normal forms" or patterns that lie beyond the evolution of populations. However, when the size of observation windows is small (for example $ws =30$, $ws =50$, and $ws =100$), the prediction performance will decrease evidently, where the values of \emph{ACC} are around $0.6$. The reasons why inaccurate predictions along with smaller observation windows and stronger selection (Figure \ref{fig:acc}(f)) can be that: (1) the increase of the collapse of cooperation at strong selection leads to balanced labels of recovery cases (i.e., \emph{AllC}) and collapse cases (i.e., \emph{AllD}), making it difficult to distinguish between each other through machine learning models; (2) small sizes of observation windows could not capture enough effective distinctive features for the sequences of evolution as inputs of the machine learning models, and the models may produce misleading and confusing predicted outcomes. 

\subsection{Higher value of temptation to defect leads to better prediction performance}

\begin{figure}[!htb]
	\setlength{\abovecaptionskip}{0pt}
	\setlength{\belowcaptionskip}{0pt}
	\centering
	\subfigure[Varying $S$s and models with fixed $T$]
	{		\includegraphics[width=0.48\linewidth]{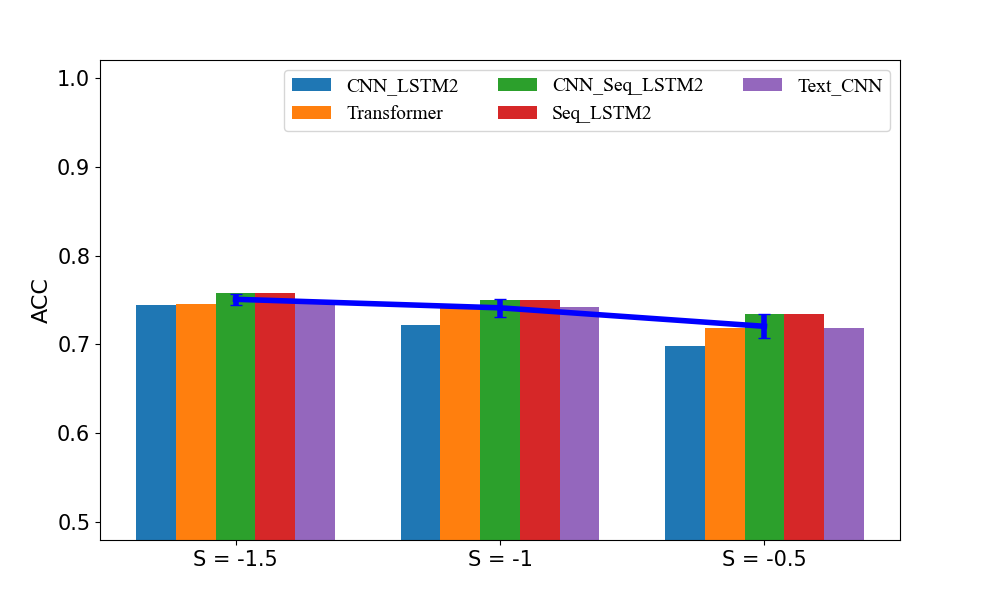}
	}
         \hskip -10pt
	\subfigure[Varying $T$s and models with fixed $S$]
	{		
            \includegraphics[width=0.48\linewidth]{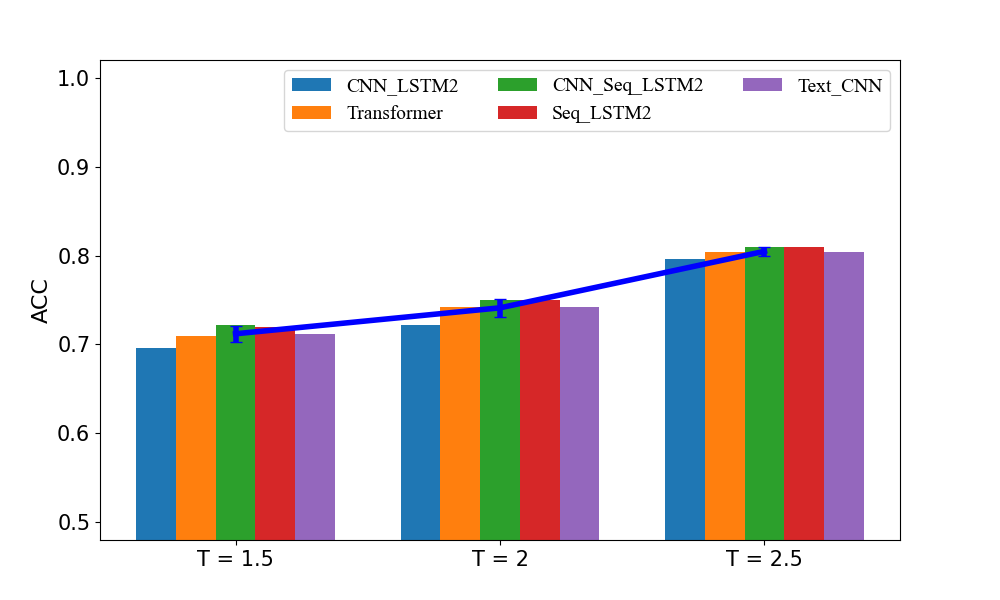}
        }	
    \subfigure[Varying $S$s and windows with fixed $T$]
	{		\includegraphics[width=0.48\linewidth]{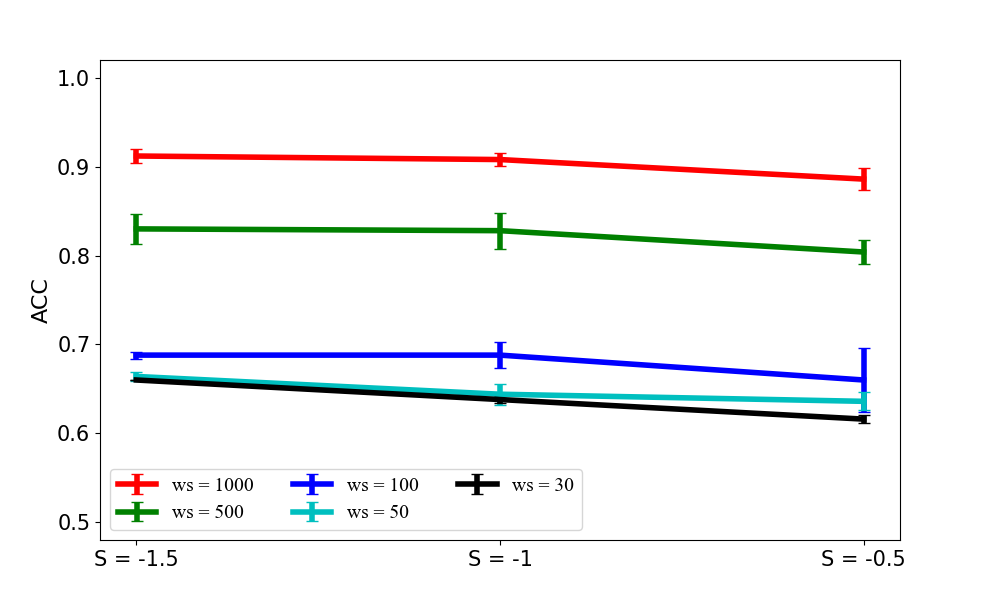}
	}
         \hskip -10pt
	\subfigure[Varying $T$s and windows with fixed $S$]
	{		
            \includegraphics[width=0.48\linewidth]{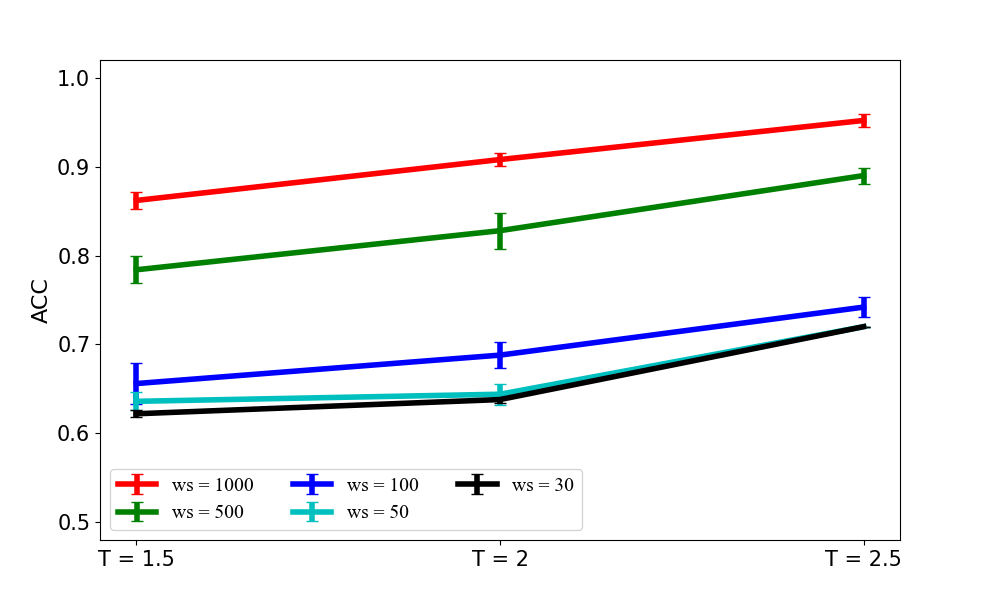}
        }	   
	\caption{The accuracy of early warning predictions is higher with the increase of game parameter $T$. For each scenario, $20000$ identical and independent simulations are conducted in a population mixed by cooperators and defectors. The game payoff matrix is set to be: $R=1$, $P=0$, and other parameters are set as before. The strength of selection $w=0.1$. We present the results by fixing $S=-1$ or $T=2$. Each case in (a)(b) is predicted by $5$ different machine learning models, and each accuracy value is an average from different window sizes. Each case in (c)(d) shows the performances of $5$ different observation windows, and each accuracy value is an average from different machine learning models. We also present the trends of average accuracy for varying $S$s and $T$s. A clear increase can be obtained  when the values of $T$ are increasing.}
	\label{fig:accST}
\end{figure}

The evolution of population is highly associated with the games played between cooperators and defectors. 
It is clear that the prediction performances are significantly dependent on observation windows as well as the strength of selection (Figure \ref{fig:acc}). As for the game parameters, the values of $S$ and $T$ are important to the performance of prediction, where the two parameters can be used to measure the level of advantage to defect. Smaller values of $S$ will bring more loss for cooperators, and higher values of $T$ will provide more benefits for defectors. In this study, we keep $R$ and $P$ fixed but vary $S$ and $T$. To reveal the influences of game parameters $S$ and $T$ on the performance of early warnings, we calculated the average values of accuracy by fixing $S$ \& varying $T$ (Figure \ref{fig:accST}(a)(c)),  and fixing $T$ \& varying $T$ (Figure \ref{fig:accST}(b)(d)), under strong selection ($w=0.1$). As we can see in Figure \ref{fig:accST}, the prediction accuracy performance is dependent on the game parameters of $S$ and $T$ as well as the machine learning models and observation windows. Interestingly, with the increase of $T$, which means the value of temptation to defect in games, the machine learning models can improve prediction performance. In addition, the models of \textbf{CNN-Seq-LSTM} and \textbf{Seq-LSTM} out-compete others in all cases, indicating the effectiveness of early warnings based on the combination of LSTM and CNN.

\subsection{Heterogeneous networks are resilient to the invasions and foster better prediction}

\begin{figure}[!htb]
	\setlength{\abovecaptionskip}{0pt}
	\setlength{\belowcaptionskip}{0pt}
	\centering
	 \subfigure[The recovery time]
	{		\includegraphics[width=0.45\linewidth]{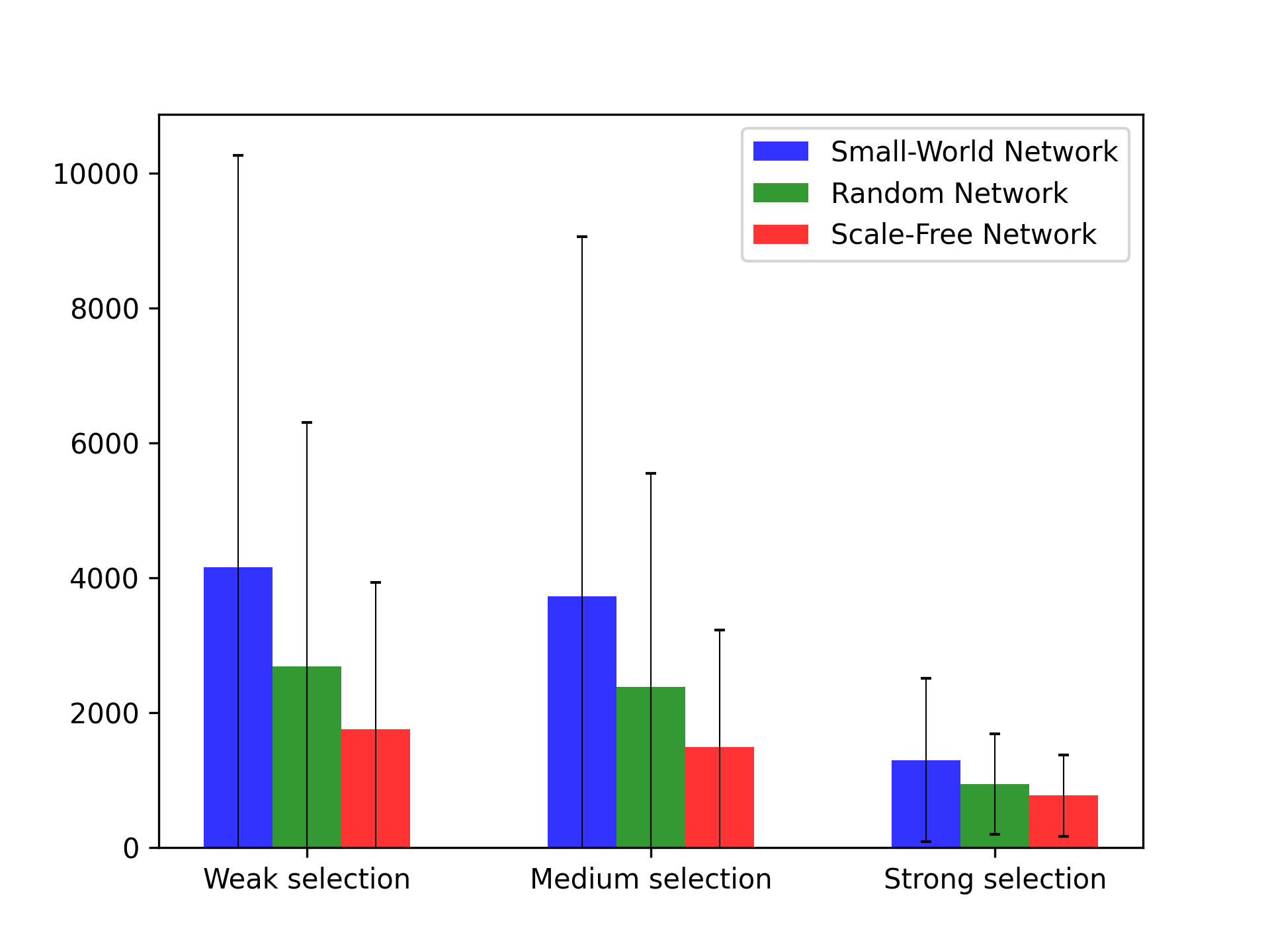}
	}
         \hskip -5pt
	\subfigure[The collapse time]
	{		
            \includegraphics[width=0.445\linewidth]{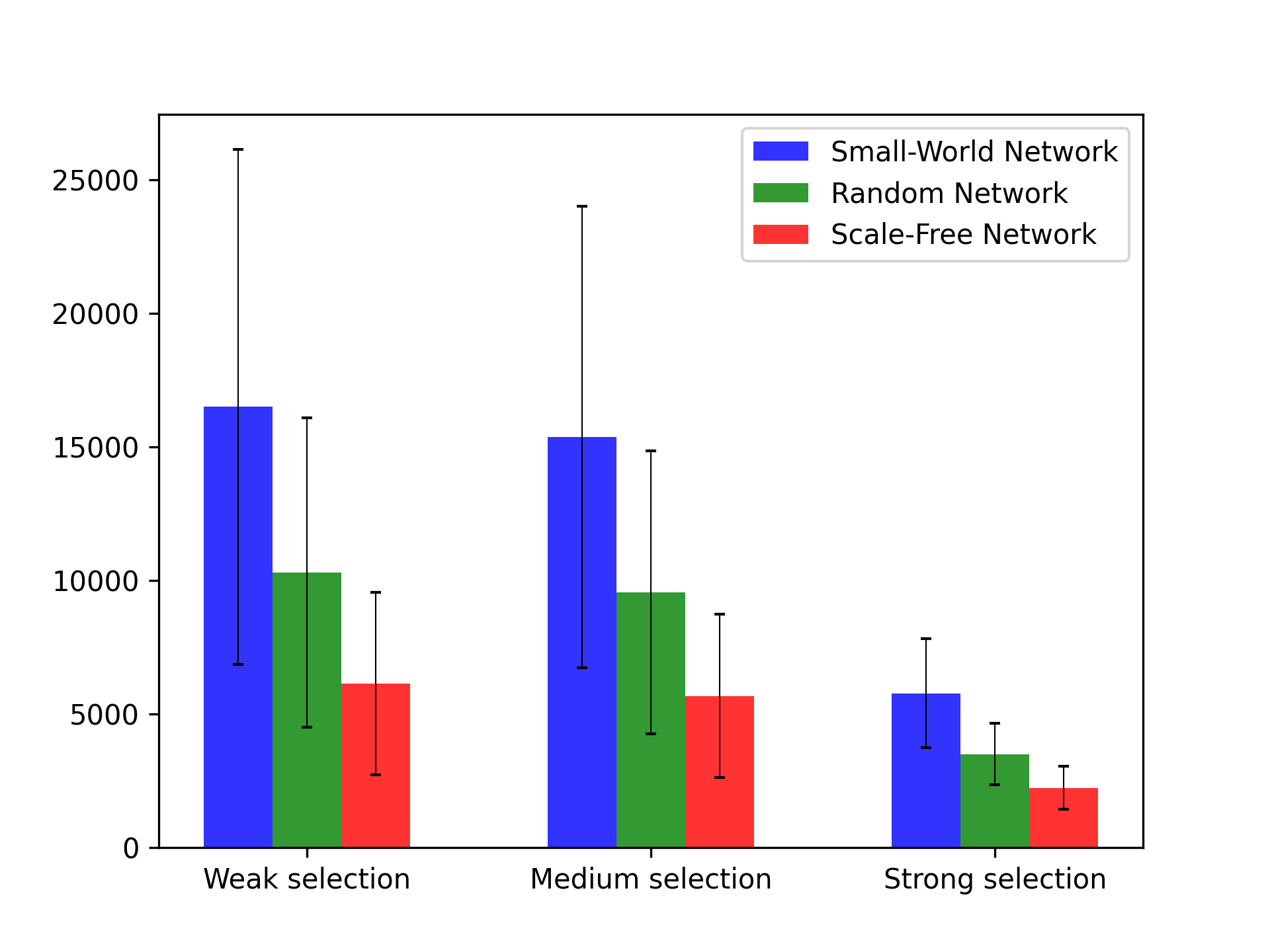}
        }	  
        \subfigure[The probability of collapse]
	{		\includegraphics[width=0.45\linewidth]{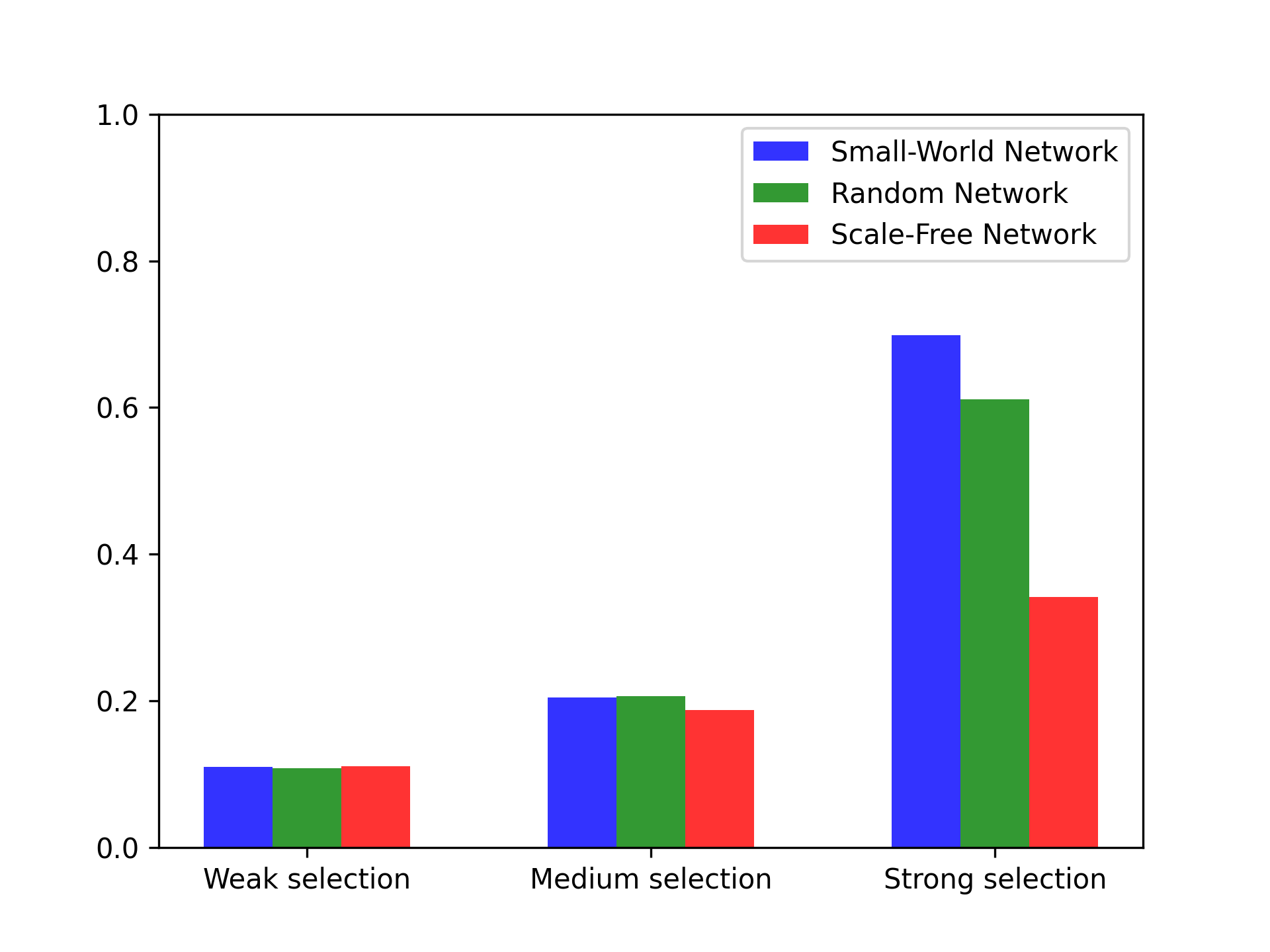}
	}
         \hskip -5pt
	\subfigure[The accuracy of early warning prediction]
	{	            \includegraphics[width=0.445\linewidth]{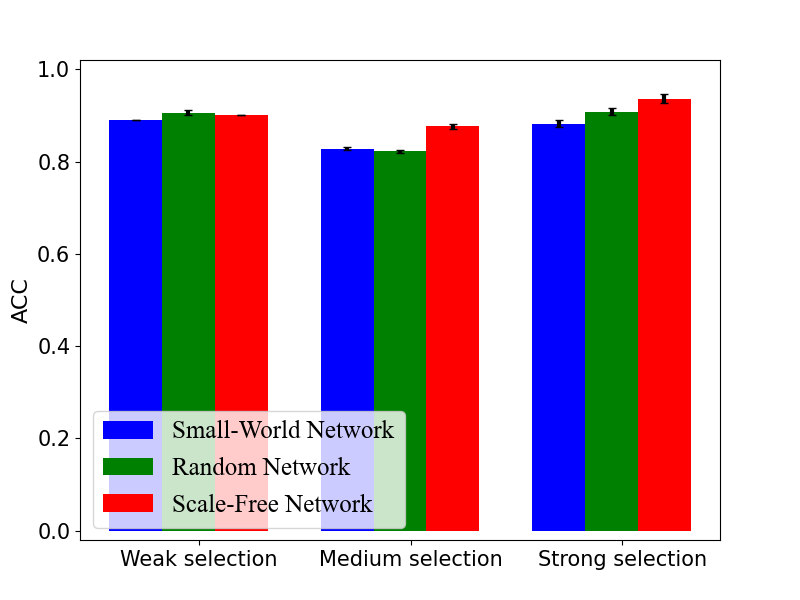}
        }	
	\caption{The evolution of cooperation and accuracy of early warning prediction is dependent on the population structures. Three types of networks are considered: small-world networks, random networks and scale-free networks. The strength of selection is set to be weak with $w=0.001$, medium with $w=0.01$, and strong with $w=0.1$. For each scenario, $20000$ identical and independent simulations are conducted in a population mixed by cooperators and defectors. The game payoff matrix is set to be: $R=1$, $S=-1$, $T=2$, $P=0$, and other parameters are set as before. The accuracy value is an average from $5$ machine learning models.}
	\label{fig:net}
\end{figure}

The structure of the community is also an important factor in the evolution of cooperation as well as the performance of early warnings. In addition to random networks, we will consider small-world networks and scale-free networks, to illustrate the influence of network heterogeneity on the evolution of cooperation and prediction performance. From small-world networks, to random networks, and then scale-free networks, the structures of networks get more heterogeneous, and the distributions of nodal degree get more diverse. As we can see, it is easier for scale-free networks to reach frozen states (either recovery or collapse of cooperation), and both recovery time (Figure \ref{fig:net}(a)) and collapse time ( Figure \ref{fig:net}(b)) decrease with the increase of selection strength. Due to the diversity of nodal degree, the clusters of cooperative component can be formed quickly \cite{yang2014older, YANG2015network}, which leads to the decrease in the probability of collapse ( Figure \ref{fig:net}(c)). More interestingly, the average  prediction accuracy in scale-free networks is higher than that in random or small-world networks (Figure \ref{fig:net}(d)), indicating the effectiveness of early warning prediction is associated with the structure of populations.

\subsection{Recovery prediction vs collapse prediction}

The outcomes of the evolutionary dynamics (recovery or collapse of cooperation) are highly dependent on the strength of selection. Due to the imbalanced samples of recovery and collapse in the training of machine learning models with the changes of selection strength, it remains a challenge to make trade-off between precision and recall for both cases. That means that one needs to train an effective model with imbalanced samples, in particular when the number of recovery or collapse samples is very different. In this case, there are huge differences between the prediction of recovery and that of collapse. To reveal the influences of selection on the prediction of recovery and collapse we analyze the differences between the positive labeling of recovery case and the positive labeling of collapse case in terms of precision, recall and F1 score.

\subsubsection{Early warning prediction in precision}

By definition,  precision measures how often machine learning models can correctly predict the recovery case or the collapse case.  For the recovery case, we need to compute the precision like that:
\begin{equation}
\begin{aligned}
       P_r  &= \frac{\#TP}{\#TP+\#FP} \\
        & = \frac{\textbf{N}(\emph{Recovery, Recovery})}{\textbf{N}(\emph{Recovery, Recovery})+\textbf{N}(\emph{Collapse, Recovery})}
\end{aligned}
\end{equation}
where $\textbf{N}(\emph{x, y})$ means the number of samples where the actual is $x$ and the predicted is $y$.

On the contrary, for the collapse case, we need to compute the precision like that:
\begin{equation}
\begin{aligned}
       P_c  &= \frac{\#TP}{\#TP+\#FP} \\
        & = \frac{\textbf{N}(\emph{Collapse, Collapse})}{\textbf{N}(\emph{Collapse, Collapse})+\textbf{N}(\emph{Recovery, Collapse})}
\end{aligned}
\end{equation}

\begin{figure}[!htb]
	\setlength{\abovecaptionskip}{0pt}
	\setlength{\belowcaptionskip}{0pt}
	\centering
	\subfigure[Weak selection]
	{		\includegraphics[width=0.90\linewidth]{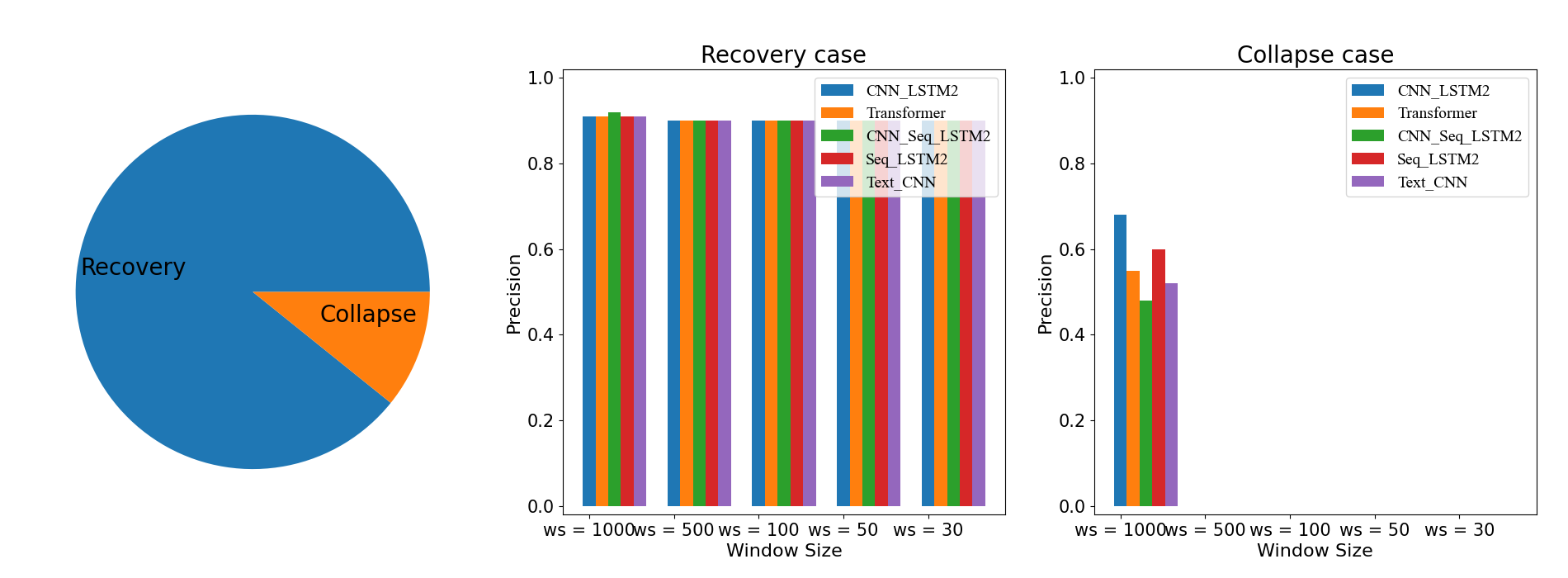}
	}
         \hskip -5pt
	\subfigure[Medium selection]
	{		\includegraphics[width=0.90\linewidth]{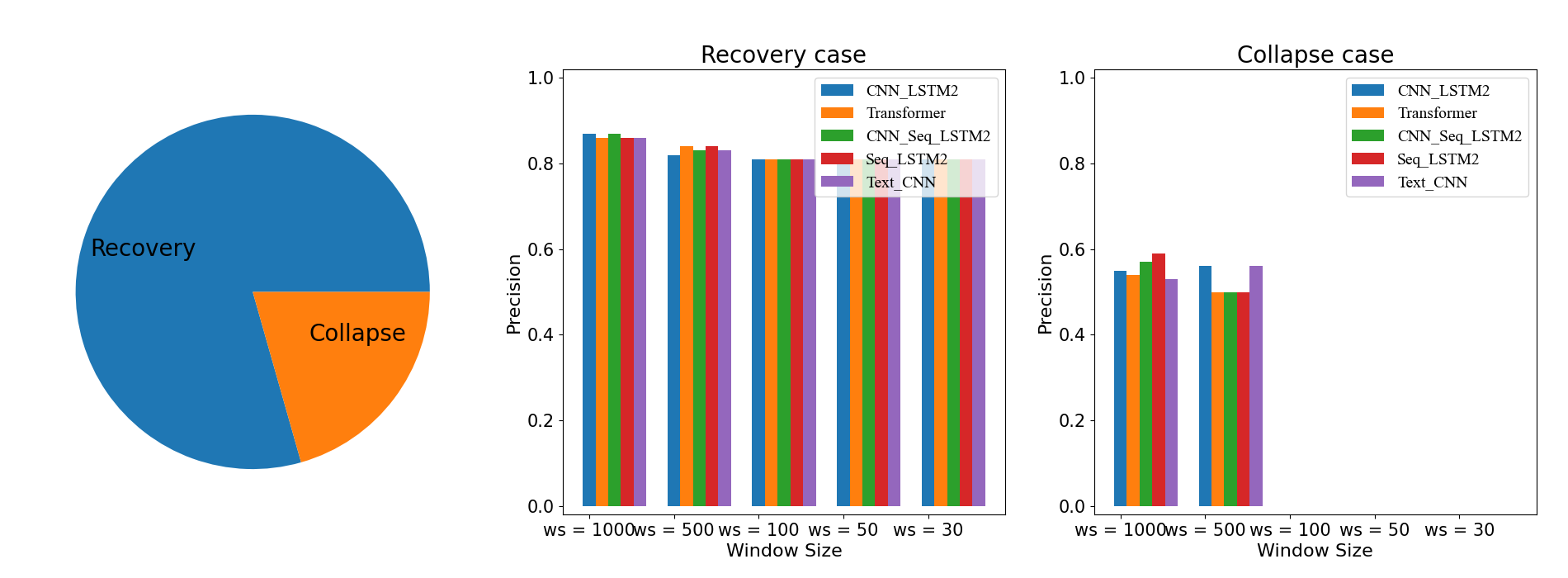}	
	}	
 \hskip -5pt
	\subfigure[Strong selection]
	{		\includegraphics[width=0.90\linewidth]{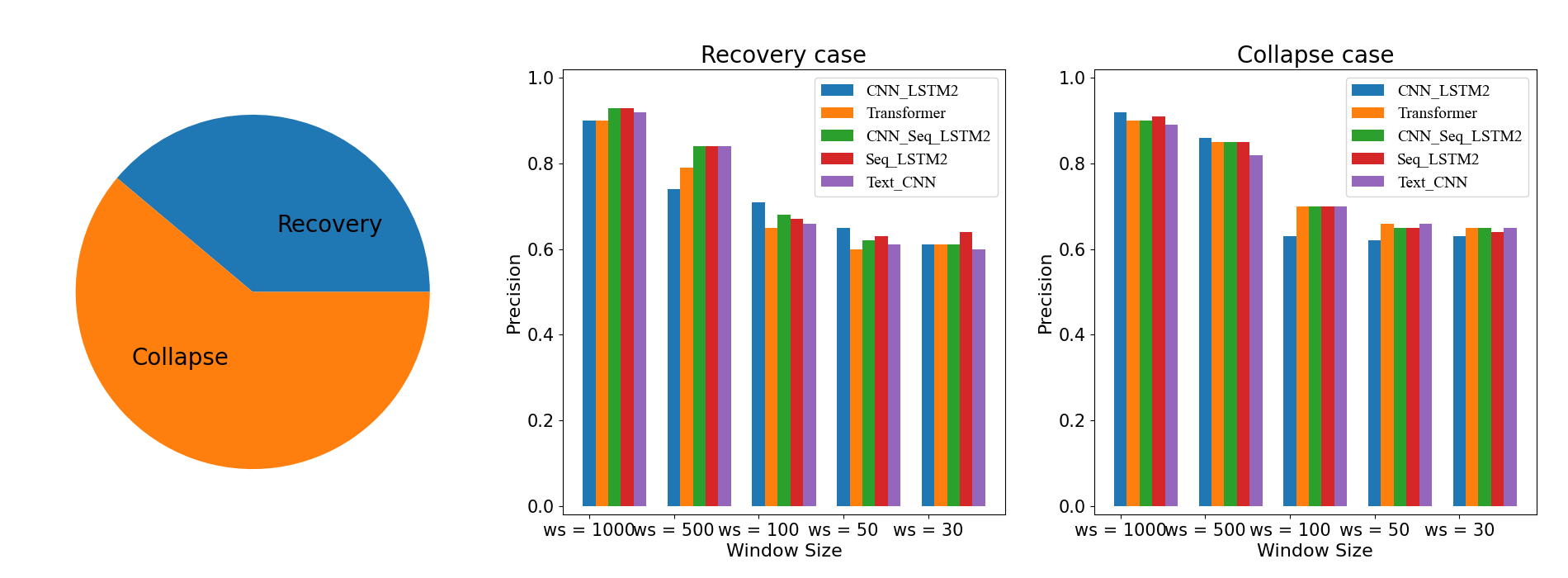}
	}	
	\caption{Precision for the recovery of cooperation and collapse of cooperation under various selections.  Early warning prediction is dependent on the machine learning models, the strength of selection and the size of observation windows. Total $20000$ identical and independent simulations are conducted in a population with $N=100$ agents, where the initial fraction of defectors is $\eta = 10\%$, and that of cooperators is $1-\eta = 90\%$. The fractions of recovery samples and collapse samples are depicted by the pie charts. The game payoff matrix is set to be: $R=1$, $P=0$, $S =-1$, and $T=2$. Representative $5$ machine learning models are trained to predict the oncoming recovery or collapse at (a) weak selection $w=0.001$,  (b) medium selection $w=0.01$  and (c) strong selection $w=0.1$.}
	\label{fig:precision}
\end{figure}

By changing the strength of selection and the size of observation windows, we present the precision from the perspective of recovery case and collapse case. Interestingly, the prediction precision of the two cases is significantly different, which is caused by the imbalanced samples in the training of machine learning models. Moreover, the values of precision is also affected by the changing windows and selections. For collapse case, the precision is quite small at weak and medium selection (Figure \ref{fig:precision}(a)(b)), especially when the window size is small, where the models could not predict the collapse of population from the available samples. For recovery case, however, the models can learn the patterns well, and the prediction is successful. Furthermore, the precision is higher for larger size of observation window at strong selection (Figure \ref{fig:precision}(c)), which indicates that the proposed machine learning models can distinguish between recovery and collapse easily based on the long-term temporal and structural evolution samples.

\subsubsection{Early warning prediction in recall}

Similar to the above definitions of precision,  for the recovery case, the recall value measures how often a machine learning model correctly identifies the recovery of cooperation from all actual recovery cases, which can be calculated by:
\begin{equation}
\begin{aligned}
       R_r  &= \frac{\#TP}{\#TP+\#FN} \\
        & = \frac{\textbf{N}(\emph{Recovery, Recovery})}{\textbf{N}(\emph{Recovery, Recovery})+\textbf{N}(\emph{ Recovery, Collapse})}
\end{aligned}
\end{equation}

On the contrary, for the collapse case, the recall value measures how often a machine learning model correctly identifies the collapse of cooperation from all actual collapse cases, which can be calculated by: 
\begin{equation}
\begin{aligned}
       R_c  &= \frac{\#TP}{\#TP+\#FN} \\
        & = \frac{\textbf{N}(\emph{Collapse, Collapse})}{\textbf{N}(\emph{Collapse, Collapse})+\textbf{N}(\emph{Collapse, Recovery})}
\end{aligned}
\end{equation}

\begin{figure}[!htb]
	\setlength{\abovecaptionskip}{0pt}
	\setlength{\belowcaptionskip}{0pt}
	\centering
	\subfigure[Weak Selection]
	{		\includegraphics[width=0.90\linewidth]{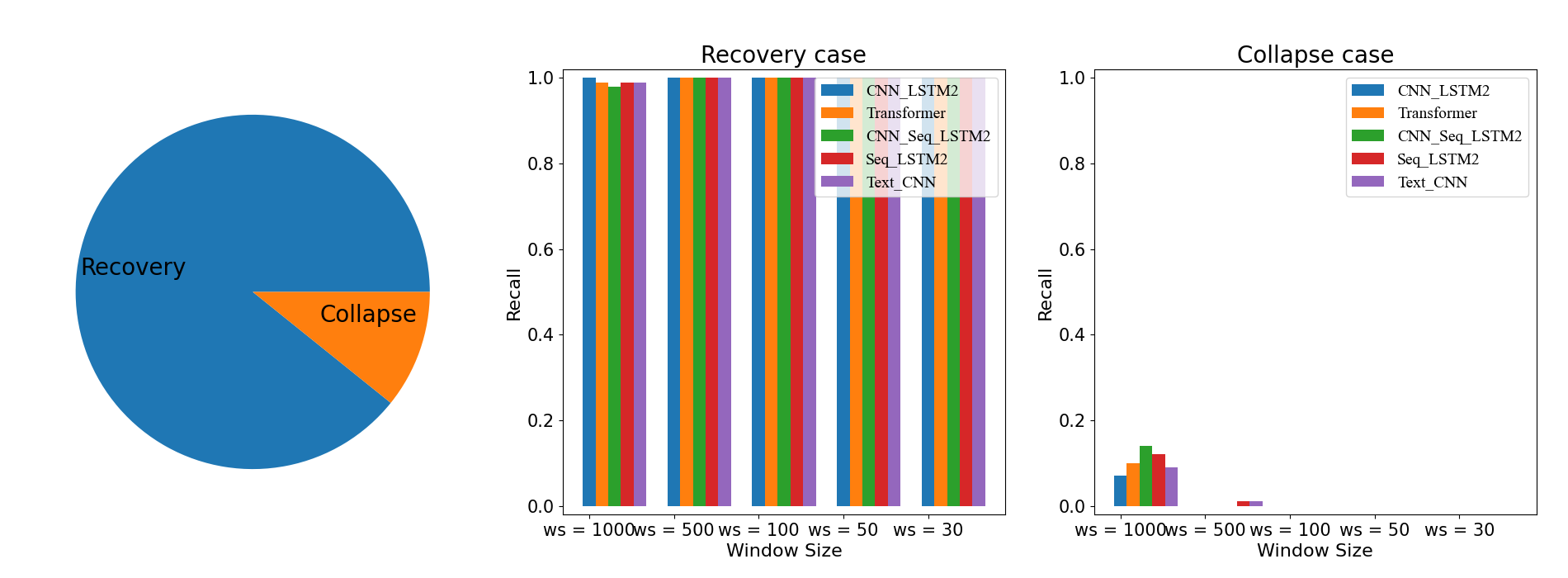}
	}
 \hskip -5pt
	\subfigure[Medium Selection]
	{		\includegraphics[width=0.90\linewidth]{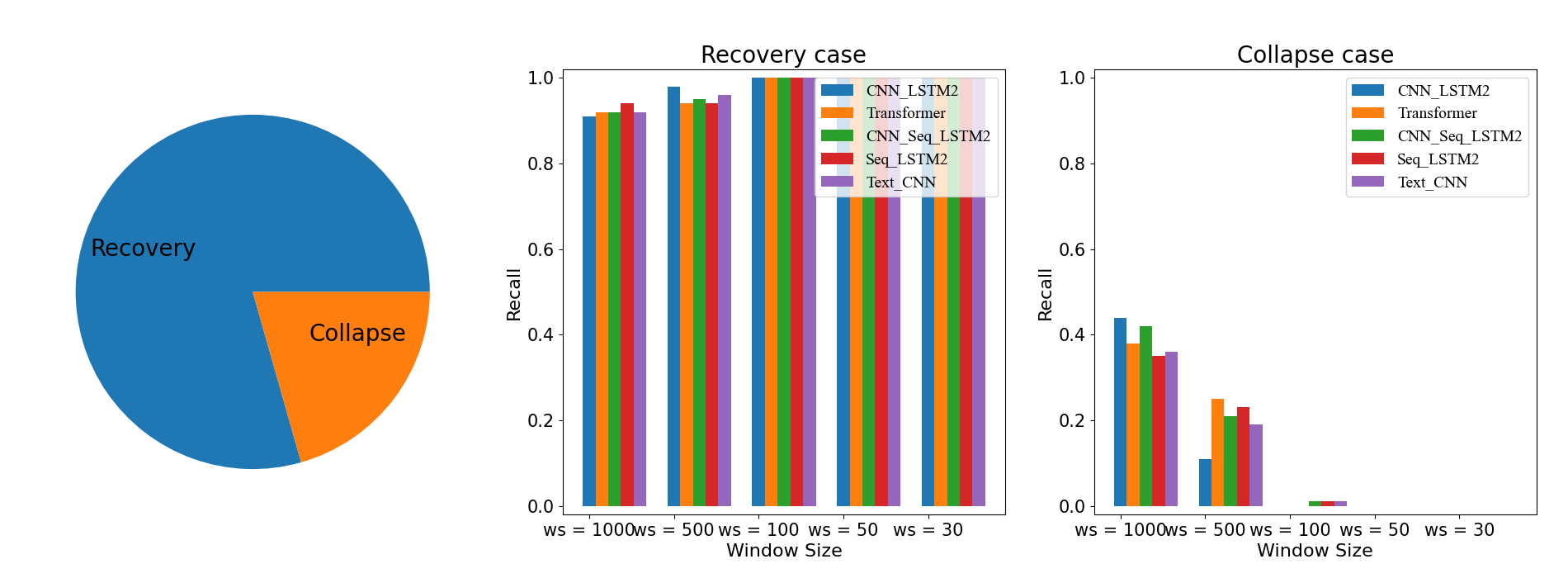}	
	}	
 \hskip -5pt
	\subfigure[Strong Selection]
	{		\includegraphics[width=0.90\linewidth]{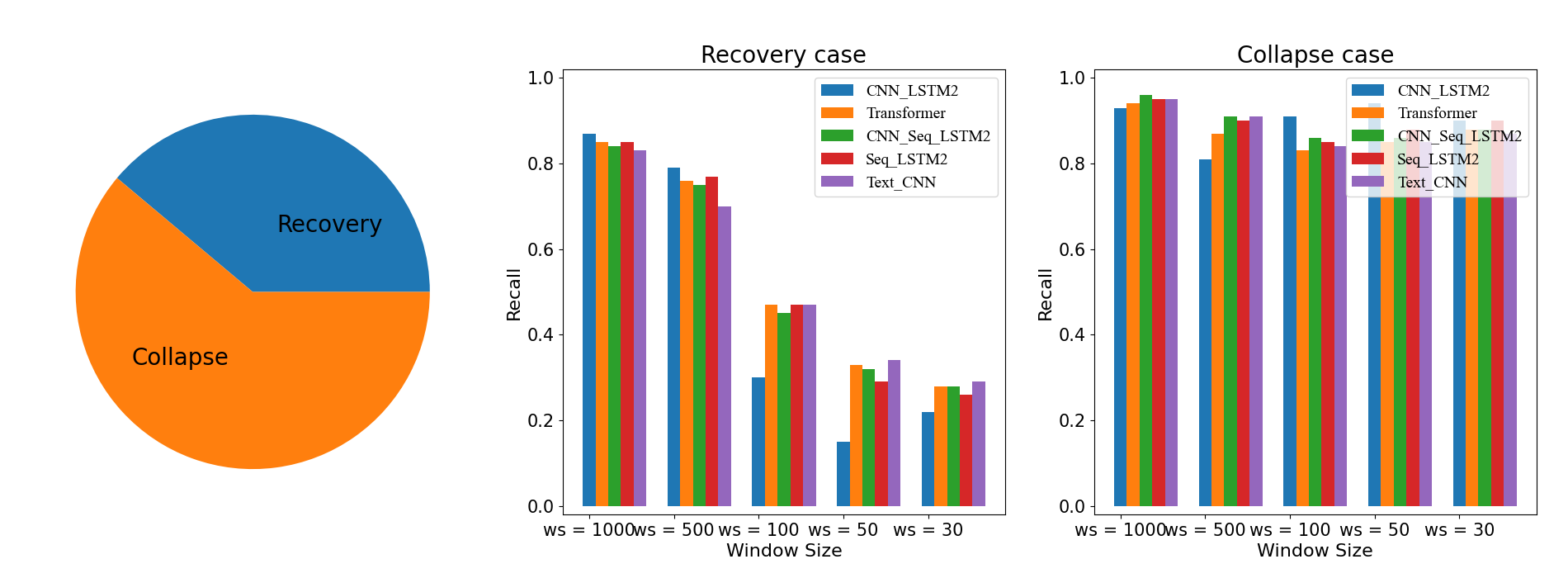}	
	}	
	\caption{High recall of early warning prediction for both recovery case and collapse case requires large observation windows and strong selection strength. Total $20000$ identical and independent simulations are conducted in a population with $N=100$ agents, where the initial fraction of defectors is $\eta = 10\%$, and that of cooperators is $1-\eta = 90\%$. The fractions of recovery samples and collapse samples are depicted by the pie charts. The game payoff matrix is set to be: $R=1$, $P=0$, $S =-1$, and $T=2$. Representative $5$ machine learning models are trained to predict the oncoming recovery or collapse at (a) weak selection $w=0.001$, (b) medium selection $w=0.01$  and (c) strong selection $w=0.1$. }
	\label{fig:recall}
\end{figure}

Clearly, a high value of $R_r$ indicates that cooperation recovery is more likely to be predicted, and a high value of $R_c$ indicates that cooperation collapse is more likely to be predicted. With the increase of selection strength $w$, the recall of collapse case is enhanced, however, that of the recovery case is decreased (Figure \ref{fig:recall}).  Notably, the machine learning models could not work well in the identification of true population collapse from all cases at weak (Figure \ref{fig:recall}(a)) and medium selection (Figure \ref{fig:recall}(b)), especially when the observation windows are small. That is caused by biased prediction for collapse from the machine learning models, which are trained by imbalanced samples. On the contrary, the recall of the recovery case is high at weak and medium selections. Only when the observation window is large and selection strength is strong (Figure \ref{fig:recall}(c)), both recovery case and collapse case can gain high recall values. However, the recall of recovery case at small observation windows is smaller than that of the collapse case, which means the collapse can be better captured by temporal and structural information at strong selection.

\subsubsection{Early warning prediction in F1 score}

The calculation of the F1 score in terms of the recovery case, can be calculated by:

\begin{figure*}[t]  
  \begin{equation}
  \begin{aligned}
         F1_r  &= \frac{\#TP}{\#TP+0.5(\#FP+\#FN)} \\
          & = \frac{\textbf{N}(\emph{Recovery, Recovery})}{\textbf{N}(\emph{Recovery, Recovery})+0.5(\textbf{N}(\emph{Collapse, Recovery})+\textbf{N}(\emph{Recovery, Collapse}))}
  \end{aligned}
  \end{equation}
\end{figure*}

For the collapse case, the F1 score is calculated by: 
\begin{figure*}[t]
\begin{equation}
\begin{aligned}
       F1_c  &= \frac{\#TP}{\#TP+0.5(\#FP+\#FN)} \\
        & = \frac{\textbf{N}(\emph{Collapse, Collapse})}{\textbf{N}(\emph{Collapse, Collapse})+0.5(\textbf{N}(\emph{Recovery, Collapse})+\textbf{N}(\emph{Collapse, Recovery}))}
\end{aligned}
\end{equation}
\end{figure*}

\begin{figure}[!htb]
	\setlength{\abovecaptionskip}{0pt}
	\setlength{\belowcaptionskip}{0pt}
	\centering
	\subfigure[Weak Selection]
	{		\includegraphics[width=0.90\linewidth]{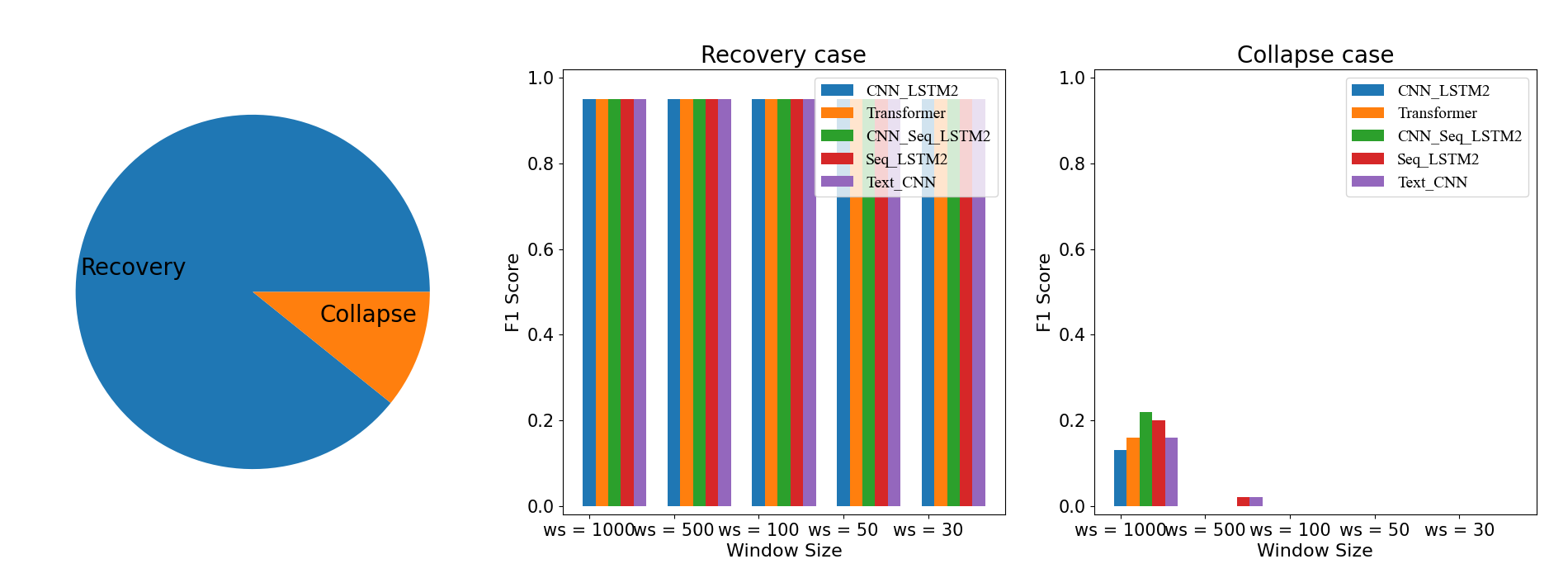}
	}
         \hskip -5pt
	\subfigure[Medium Selection]
	{		\includegraphics[width=0.90\linewidth]{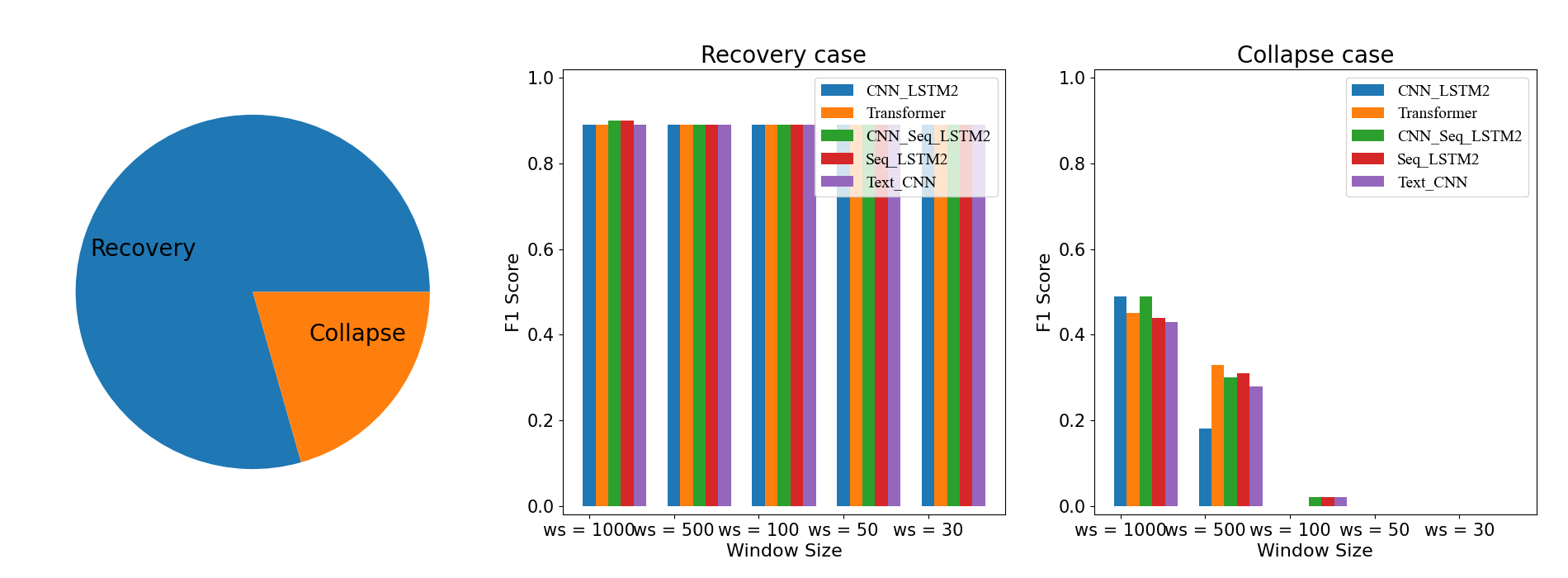}	
	}	
 \hskip -5pt
	\subfigure[Strong Selection]
	{		\includegraphics[width=0.90\linewidth]{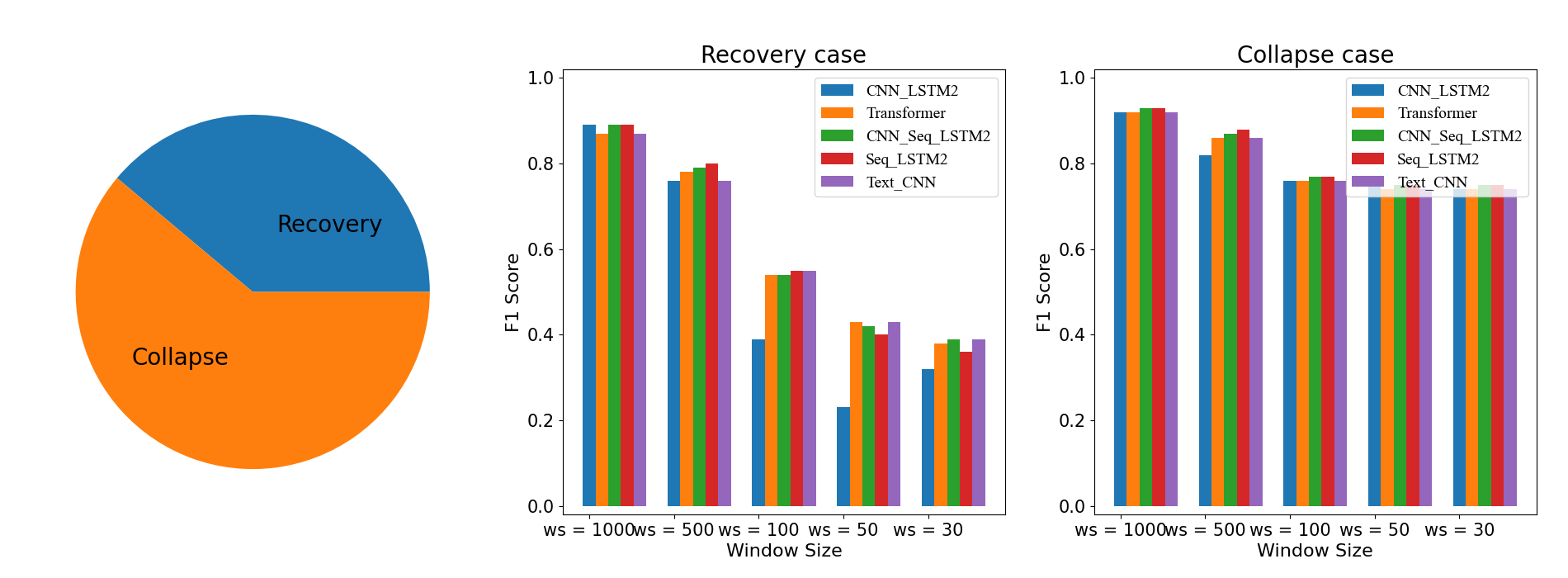}	
	}	
	\caption{F1 scores of recovery case and collapse case are dependent on the machine learning models, selection strengths, observation windows. Total $20000$ identical and independent simulations are conducted in a population with $N=100$ agents, where the initial fraction of defectors is $\eta = 10\%$, and that of cooperators is $1-\eta = 90\%$.  The fractions of recovery samples and collapse samples are depicted by the pie charts. The game payoff matrix is set to be: $R=1$, $P=0$, $S =-1$, and $T=2$. Representative $5$ machine learning models are trained to predict the oncoming recovery or collapse at (a) weak selection $w=0.001$,  (b) medium selection $w=0.01$  and (c) strong selection $w=0.1$.}
	\label{fig:f1score}
\end{figure}

As we can see in Figure \ref{fig:f1score}, the prediction performance of F1 for the collapse case is enhanced with the increase of selection strength. For the recovery case, however, the performance of F1 decreases when selection is strong, especially when the observation windows are small at strong selection (Figure \ref{fig:f1score}(c)). These results reveal that the prediction of cooperation collapse through machine learning models is effective under strong selections, while the recovery of cooperation is better predicted under weak selections.

\section{Conclusions}
\label{conclusion}

How to predict the future failure or success of structured communities is never an easy job, particularly when the dynamics of the population is based on a stochastic evolutionary dynamics and the fate of the community is based on the outcome of the conflict between cooperators and cheaters.
In this paper, we investigate early warnings for the evolutionary dynamics of an organized population using machine learning with temporal and structural data.  Using evolutionary graph theory, we develop a methodology to predict whether, after a small initial cheater invasion, the population will eliminate the invaders and reach a state of all cooperators (i.e., the recovery of cooperation), or the state of all defectors (i.e., the collapse of cooperation). The methodology developed uses machine learning and temporal and structural information within an observation window. 
As the dynamics of evolutionary populations is generally controlled by the game parameters, the strength of selection and population structure, we constructed several machine learning models, including Seq-LSTM, CNN-Seq-LSTM, CNN-LSTM, Text-CNN, Transformer, etc, to illustrate the prediction performance under different regimes and different observation window size.
Interestingly, we find that these machine learning models are indifferent at weak selection, where the probability of collapse of cooperation is generally quite low. 
However, when the strength of selection is strong, only sufficiently large observation windows can lead to accurate early warning predictions. In that case, the models of CNN-Seq-LSTM and Seq-LSTM out-compete the others. 
In general, we observe that the typology of game played between cooperators and cheaters is particularly important: larger advantage to cheat is associated with a population dynamics that can be predicted with better accuracy. Moreover, the organization of the population is important too: predictions are generally more accurate when the population is organized on scale-free networks.
Furthermore, for the recovery of cooperation and collapse of cooperation prediction, the metrics of precision, recall and F1 score, perform quite differently - the recovery of cooperation can be better predicted at weak selection, while the collapse of cooperation can be better predicted at strong selection.   
Overall, this work introduces a machine learning approach into detecting abrupt shifts in evolutionary graph theory and offers potential strategies for anticipating and preventing cooperation collapse in complex networks.

More generally, we believe that, given the generality and wide applicability of evolutionary graph theory,  this initial study highlights the opportunities and challenges of the interplay between population organization, game-theoretical interactions and machine learning to forecast the dynamics of complex structured populations.

\bibliographystyle{unsrt}
\bibliography{ref.bib}

\begin{thebibliography}{10}

\bibitem{Scheffer2009early}
Marten Scheffer, Jordi Bascompte, William~A. Brock, Victor Brovkin, Stephen~R. Carpenter, Vasilis Dakos, Hermann Held, Egbert~H. van Nes, Max Rietkerk, and George Sugihara.
\newblock Early-warning signals for critical transitions.
\newblock {\em Nature}, 461(7260):53--59, 2009.

\bibitem{esd-15-1117-2024}
V.~Dakos, C.~A. Boulton, J.~E. Buxton, J.~F. Abrams, B.~Arellano-Nava, D.~I. Armstrong~McKay, S.~Bathiany, L.~Blaschke, N.~Boers, D.~Dylewsky, C.~L\'opez-Mart\'{\i}nez, I.~Parry, P.~Ritchie, B.~van~der Bolt, L.~van~der Laan, E.~Weinans, and S.~K\'efi.
\newblock Tipping point detection and early warnings in climate, ecological, and human systems.
\newblock {\em Earth System Dynamics}, 15(4):1117--1135, 2024.

\bibitem{Dakos2019eco}
Vasilis Dakos, Blake Matthews, Andrew~P. Hendry, Jonathan Levine, Nicolas Loeuille, Jon Norberg, Patrik Nosil, Marten Scheffer, and Luc De~Meester.
\newblock Ecosystem tipping points in an evolving world.
\newblock {\em Nature Ecology \& Evolution}, 3(3):355-- 362, 2019.

\bibitem{Strogatz2015nonlinear}
Steven~H. Strogatz.
\newblock {\em Nonlinear Dynamics and Chaos: With Applications to Physics, Biology, Chemistry, and Engineering}.
\newblock CRC Press, 2015.

\bibitem{li2023non}
Li~Xu, Denis Patterson, Simon~Asher Levin, and Jin Wang.
\newblock Non-equilibrium early-warning signals for critical transitions in ecological systems.
\newblock {\em Proceedings of the National Academy of Sciences}, 120(5):e2218663120, 2023.

\bibitem{PELISSIE2024110429}
Mathieu Pélissié, Vincent Devictor, and Vasilis Dakos.
\newblock A systematic approach for detecting abrupt shifts in ecological timeseries.
\newblock {\em Biological Conservation}, 290:110429, 2024.

\bibitem{dakos2015resilience}
Vasilis Dakos, Stephen~R Carpenter, Egbert~H van Nes, and Marten Scheffer.
\newblock Resilience indicators: prospects and limitations for early warnings of regime shifts.
\newblock {\em Philosophical Transactions of the Royal Society B: Biological Sciences}, 370(1659):20130263, 2015.

\bibitem{Dylewsky2023evo}
Dylewsky Daniel, Lenton~Timothy M., Scheffer Marten, Bury~Thomas M., Fletcher~Christopher G., Anand Madhur, and Bauch~Chris T.
\newblock Universal early warning signals of phase transitions in climate systems.
\newblock {\em J R Soc Interface}, 20:20220562, 2023.

\bibitem{Thomas2021deep}
Thomas~M. Bury, R.~I. Sujith, Induja Pavithran, Marten Scheffer, Timothy~M. Lenton, Madhur Anand, and Chris~T. Bauch.
\newblock Deep learning for early warning signals of tipping points.
\newblock {\em Proceedings of the National Academy of Sciences}, 118(39):e2106140118, 2021.

\bibitem{Drake2020Alternative}
John~M. Drake, Suzanne~M. O’Regan, Vasilis Dakos, Sonia Kéfi, and Pejman Rohani.
\newblock Alternative stable states, tipping points, and early warning signals of ecological transitions.
\newblock In {\em Theoretical Ecology: concepts and applications}. Oxford University Press, 05 2020.

\bibitem{axelrod1981evolution}
Robert Axelrod and William~D Hamilton.
\newblock The evolution of cooperation.
\newblock {\em Science}, 211(4489):1390--1396, 1981.

\bibitem{sachs2004evolution}
Joel~L Sachs, Ulrich~G Mueller, Thomas~P Wilcox, and James~J Bull.
\newblock The evolution of cooperation.
\newblock {\em The Quarterly review of biology}, 79(2):135--160, 2004.

\bibitem{hammerstein2003genetic}
Peter Hammerstein et~al.
\newblock {\em Genetic and cultural evolution of cooperation}.
\newblock MIT press, 2003.

\bibitem{WU2023133692}
Yu’e Wu, Guoli Yang, Yu~Li, Zhipeng Zhang, Jingjing Li, and Shuhua Zhang.
\newblock Evolution of cooperation in multigames on interdependent networks.
\newblock {\em Physica D: Nonlinear Phenomena}, 447:133692, 2023.

\bibitem{nowak_s06}
M.~A. Nowak.
\newblock Five rules for the evolution of cooperation.
\newblock {\em Science}, 314:1560--1563, 2006.

\bibitem{wang2025graph}
Mengzhu Wang, Houcheng Su, Sijia Wang, Shanshan Wang, Nan Yin, Li~Shen, Long Lan, Liang Yang, and Xiaochun Cao.
\newblock Graph convolutional mixture-of-experts learner network for long-tailed domain generalization.
\newblock {\em IEEE Transactions on Circuits and Systems for Video Technology}, 2025.

\bibitem{cavaliere2012prosperity}
Matteo Cavaliere, Sean Sedwards, Corina~E Tarnita, Martin~A Nowak, and Attila Csik{\'a}sz-Nagy.
\newblock Prosperity is associated with instability in dynamical networks.
\newblock {\em Journal of theoretical biology}, 299:126--138, 2012.

\bibitem{levin2010crossing}
Simon Levin.
\newblock Crossing scales, crossing disciplines: collective motion and collective action in the global commons.
\newblock {\em Philosophical Transactions of the Royal Society B: Biological Sciences}, 365(1537):13--18, 2010.

\bibitem{cavaliere2016detecting}
Matteo Cavaliere, Guoli Yang, Vincent Danos, and Vasilis Dakos.
\newblock Detecting the collapse of cooperation in evolving networks.
\newblock {\em Scientific reports}, 6(1):30845, 2016.

\bibitem{lieberman2005evolutionary}
Erez Lieberman, Christoph Hauert, and Martin~A Nowak.
\newblock Evolutionary dynamics on graphs.
\newblock {\em Nature}, 433(7023):312--316, 2005.

\bibitem{perc2010coevolutionary}
Matja{\v{z}} Perc and Attila Szolnoki.
\newblock Coevolutionary games—a mini review.
\newblock {\em BioSystems}, 99(2):109--125, 2010.

\bibitem{Ohtsuki2006}
Hisashi Ohtsuki, Christoph Hauert, Erez Lieberman, and Martin~A. Nowak.
\newblock A simple rule for the evolution of cooperation on graphs.
\newblock {\em Nature}, 441(7092):502--505, 2006.

\bibitem{kuo2024evolutionary}
Yang~Ping Kuo and Oana Carja.
\newblock Evolutionary graph theory beyond pairwise interactions: higher-order network motifs shape times to fixation in structured populations.
\newblock {\em PLOS Computational Biology}, 20(3):e1011905, 2024.

\bibitem{shakarian2012review}
Paulo Shakarian, Patrick Roos, and Anthony Johnson.
\newblock A review of evolutionary graph theory with applications to game theory.
\newblock {\em Biosystems}, 107(2):66--80, 2012.

\bibitem{ohtsuki2007evolutionary}
Hisashi Ohtsuki, Jorge~M Pacheco, and Martin~A Nowak.
\newblock Evolutionary graph theory: Breaking the symmetry between interaction and replacement.
\newblock {\em Journal of Theoretical Biology}, 246(4):681--694, 2007.

\bibitem{sharma2023self}
Nikhil Sharma, Sedigheh Yagoobi, and Arne Traulsen.
\newblock Self-loops in evolutionary graph theory: Friends or foes?
\newblock {\em PLoS Computational Biology}, 19(9):e1011387, 2023.

\bibitem{wang2024smooth}
Mengzhu Wang, Junze Liu, Ge~Luo, Shanshan Wang, Wei Wang, Long Lan, Ye~Wang, and Feiping Nie.
\newblock Smooth-guided implicit data augmentation for domain generalization.
\newblock {\em IEEE Transactions on Neural Networks and Learning Systems}, 2024.

\bibitem{wang2024equity}
Mengzhu Wang, Shanshan Wang, Xun Yang, Jianlong Yuan, and Wenju Zhang.
\newblock Equity in unsupervised domain adaptation by nuclear norm maximization.
\newblock {\em IEEE Transactions on Circuits and Systems for Video Technology}, 34(7):5533--5545, 2024.

\bibitem{wang2024inter}
Mengzhu Wang, Yuehua Liu, Jianlong Yuan, Shanshan Wang, Zhibin Wang, and Wei Wang.
\newblock Inter-class and inter-domain semantic augmentation for domain generalization.
\newblock {\em IEEE Transactions on Image Processing}, 33:1338--1347, 2024.

\bibitem{alpaydin2020introduction}
Ethem Alpaydin.
\newblock {\em Introduction to machine learning}.
\newblock MIT press, 2020.

\bibitem{nowak2004evolutionary}
Martin~A Nowak and Karl Sigmund.
\newblock Evolutionary dynamics of biological games.
\newblock {\em science}, 303(5659):793--799, 2004.

\bibitem{Yang2021Strategically}
Guoli Yang, Matteo Cavaliere, Cheng Zhu, and Matjaž Perc.
\newblock Strategically positioning cooperators can facilitate the contagion of cooperation.
\newblock {\em Scientific Reports}, 11(1), 2021.

\bibitem{Yang2020Ranking}
Guoli Yang, Matteo Cavaliere, Cheng Zhu, and Matjaž Perc.
\newblock Ranking the invasions of cheaters in structured populations.
\newblock {\em Scientific Reports}, 10(1), 2020.

\bibitem{Yang2019Identification}
Guoli Yang, Tina~P. Benko, Matteo Cavaliere, Jincai Huang, and Matjaž Perc.
\newblock Identification of influential invaders in evolutionary populations.
\newblock {\em Scientific Reports}, 9(1), 2019.

\bibitem{LSTM}
Sepp Hochreiter and Jürgen Schmidhuber.
\newblock Long short-term memory.
\newblock {\em Neural Computation}, 9(8):1735--1780, 1997.

\bibitem{CNN}
Y.~Lecun, L.~Bottou, Y.~Bengio, and P.~Haffner.
\newblock Gradient-based learning applied to document recognition.
\newblock {\em Proceedings of the IEEE}, 86(11):2278--2324, 1998.

\bibitem{Transformer}
Ashish Vaswani, Noam~M. Shazeer, Niki Parmar, Jakob Uszkoreit, Llion Jones, Aidan~N. Gomez, Lukasz Kaiser, and Illia Polosukhin.
\newblock Attention is all you need.
\newblock In {\em Neural Information Processing Systems}, 2017.

\bibitem{yang2014older}
Guoli Yang, Jincai Huang, and Weiming Zhang.
\newblock Older partner selection promotes the prevalence of cooperation in evolutionary games.
\newblock {\em Journal of theoretical biology}, 359:171--183, 2014.

\bibitem{YANG2015network}
Guoli Yang, Weiming Zhang, and Baoxin Xiu.
\newblock Neighbourhood reaction in the evolution of cooperation.
\newblock {\em Journal of Theoretical Biology}, 372:118--127, 2015.

\end{thebibliography}

\end{document}